\documentstyle[epsfig,natbib2,natbibmnfix]{mn}

\newcommand{\kms}{\mbox{ km s$^{-1}$}}
\newcommand{\be}{\begin{equation}}
\newcommand{\ee}{\end{equation}}

\def\ltsima{$\; \buildrel < \over \sim \;$}
\def\simlt{\lower.5ex\hbox{\ltsima}}
\def\gtsima{$\; \buildrel > \over \sim \;$}
\def\simgt{\lower.5ex\hbox{\gtsima}}

\title{Modeling feedback from stars and black holes in galaxy mergers}

\author[V.~Springel, T.~Di~Matteo, and L.~Hernquist] {\parbox{18cm}{Volker
Springel$^{1}$\footnotemark[1], Tiziana Di Matteo$^{1}$\footnotemark[2] and Lars
Hernquist$^2$\footnotemark[3]}\vspace{0.3cm}\\ $^1$Max-Planck-Institut
f\"{u}r Astrophysik, Karl-Schwarzschild-Stra\ss{}e 1, 85740 Garching
bei M\"{u}nchen, Germany\\ $^2$Harvard-Smithsonian Center for
Astrophysics, 60 Garden Street, Cambridge, MA 02138, USA}

\begin{document}

\maketitle
\begin{abstract}
We describe techniques for incorporating feedback from star formation
and black hole accretion into simulations of isolated and merging
galaxies. At present, the details of these processes cannot be
resolved in simulations on galactic scales. Our basic approach
therefore involves forming coarse-grained representations of the
properties of the interstellar medium and black hole accretion
starting from basic physical assumptions, so that the impact of these
effects can be included on resolved scales. We illustrate our method
using a multiphase description of star-forming gas. Feedback from star
formation pressurises highly overdense gas, altering its effective
equation of state. We show that this allows the construction of stable
galaxy models with much larger gas fractions than possible in earlier
numerical work. We extend the model by including a treatment of gas
accretion onto central supermassive black holes in galaxies.  Assuming
thermal coupling of a small fraction of the bolometric luminosity of
accreting black holes to the surrounding gas, we show how this
feedback regulates the growth of black holes.  In gas-rich mergers of
galaxies, we observe a complex interplay between starbursts and
central AGN activity when the tidal interaction triggers intense
nuclear inflows of gas. Once an accreting supermassive black hole has
grown to a critical size, feedback terminates its further growth, and
expels gas from the central region in a powerful quasar-driven
wind. Our simulation methodology is therefore able to address the
coupled processes of gas dynamics, star formation, and black hole
accretion during the formation of galaxies.
\end{abstract}


\section{Introduction}
\label{intro}

\renewcommand{\thefootnote}{\fnsymbol{footnote}}
\footnotetext[1]{E-mail: volker@mpa-garching.mpg.de}
\footnotetext[2]{E-mail: tiziana@mpa-garching.mpg.de}
\footnotetext[3]{E-mail: lars@cfa.harvard.edu}

It is now recognised that galaxy collisions and mergers are relevant
to a wide range of phenomena associated with both ordinary and active
galaxies. The collisionless dynamics of this process is relatively
well-understood. The seminal work of \citet{Toomre1972} using
restricted N-body methods demonstrated that gravitational tidal forces
between disks interacting transiently can account for the narrow tails
and bridges commonly seen in morphologically peculiar galaxies.
Subsequent work using self-consistent models \citep{Barnes1988,
Barnes1992, Hernquist1992, Hernquist1993a} verified these conclusions
and made it possible to extend the calculations to investigate mergers
and the detailed structure of the remnants left behind.  These studies
showed that in many respects the remnants closely resemble elliptical
galaxies, as proposed by \citet{Toomre1977}.

However, it is clear that pure stellar dynamics is not adequate to
explain the extreme variety of objects linked to galaxy encounters.
This possibility was already anticipated by \citet{Toomre1972} who
suggested that collisions could ``bring {\it deep} into a galaxy a
fairly {\it sudden} supply of fresh fuel,'' leading to a period of
violent, enhanced star formation. The significance of
non-gravitational processes in galaxy interactions was confirmed in
the 1980s when it was recognised that the brightest infrared sources
seen with the IRAS satellite are invariably associated with peculiar
galaxies \citep{Sanders1988, Melnick1990}.  A natural interpretation
of these systems is that the infrared emission is powered by a
starburst, triggered through mergers.

There are also numerous studies indicating that quasars, radio
galaxies, and active galactic nuclei (AGN) are preferentially found in
tidally disturbed objects \citep[for reviews, see
e.g.][]{BarnesHernquist1992,Jogee2004}.  Indeed, much circumstantial
evidence suggests an evolutionary connection between merger-induced
starbursts and quasar activity \citep{Sanders1988}.  In addition, in
recent years, strong dynamical evidence has been accumulated
indicating that supermassive black holes reside at the centre of most
galaxies \citep{KormendyRichstone1995, Magorrian1998,
Kormendy2001}. Moreover, a remarkable connection between
supermassive black holes and the properties of their host galaxies has
been discovered: the masses of central supermassive black holes are
tightly correlated with the stellar velocity dispersion of their host
galaxy bulges \citep{Ferrarese2000, Gebhardt2000, Tremaine2002}, as
well as with the mass of the spheroidal component
\citep{Magorrian1998, McLure2002, Marconi2003}. This strong link
suggests a fundamental connection between the growth of black holes
and the formation of stellar spheroids in galaxy halos \citep[and
references therein]{Kauffmann2000, Volonteri2003, Wyithe2003,
Granato2004, Haiman2004, DiMatteo2003, DiMatteo2004a}.

A better understanding of galaxy interactions and mergers will,
therefore, require simultaneous accounting of the physics of
interstellar gas, star formation, the growth of supermassive black
holes in galactic nuclei, and various forms of feedback associated
with both massive stars and AGN.  As we discuss in detail below, it is
not immediately obvious how these effects should be incorporated into
simulations, primarily because we presently lack a sufficiently
developed theory of star formation, but also because the current
generation of computer models cannot resolve the complex structure of
star-forming gas on the scales of whole galaxies.  For these reasons,
efforts to study non-gravitational processes in galaxy mergers have to
rely on strong simplifications, which have often been very restrictive
in previous work.

\citet{Negroponte1983} used a sticky-particle treatment to show that
galaxy mergers can concentrate gas in the inner regions of a remnant.
Later, \citet{Hernquist1989b} and \citet{BarnesHernquist1991,
BarnesHernquist1996} examined the fate of gas in minor and major
mergers with a smoothed particle hydrodynamics (SPH) algorithm,
neglecting star formation and feedback and approximating the structure
of the interstellar medium (ISM) with an isothermal equation of state.
The first serious attempts to model star formation in galaxy mergers
were made by \citet{Hernquist1995} and \citet{Mihos1996}, who also
used an SPH method with an isothermal gas and included a minimal form
of kinetic feedback from massive stars \citep{Mihos1994a}.
\citet{Gerritsen1997} and \citet{Springel2000} began to generalise
these calculations by implementing simple approaches to allow for
departures from an isothermal equation of state.

While the earlier calculations yielded many successes by providing an
explicit theoretical link between galaxy mergers and starbursts, many
related phenomena remain poorly understood, partly because of the
approximations made in handling the consequences of feedback.  In this
paper, we introduce a new methodology to incorporate star formation
and black hole growth into galaxy-scale simulations using a
sub-resolution approach.  Starting from a ``microscopic'' physical
theory for e.g.~the ISM, we form a ``macroscopic'' coarse-grained
representation that captures the impact of star formation and black
hole growth on resolved scales.  This strategy is not tied to a
particular underlying model and can be used to graft any
well-specified theoretical description onto the simulations.  As a
specific example, in what follows we will illustrate our technique
using a sub-resolution multiphase model for star-forming gas developed
by \citet{Springel2003a}. As an additional component, we will add a
treatment of gas accretion onto supermassive black holes, modelled as
collisionless `sink' particles, and we treat feedback processes
associated with the accretion.

Note that the use of a sub-resolution approach to include unresolved
effects in simulations is a general concept also used, for example, in
N-body representations of collisionless fluids, where six-dimensional
phase space is discretised into fluid elements that are
computationally realised as particles.  Our treatment is analogous to
this, except that we endow the particles with internal degrees of
freedom that characterise, e.g., the thermodynamic state of
star-forming gas.  These internal degrees of freedom evolve in a
manner consistent with the basic theory from which they are derived,
enabling us to couple the ``microphysics'' of the ISM to the larger
scale dynamics of the galaxies.

This paper is organised as follows.  We first discuss the construction
of compound galaxy models used in our study in Section~\ref{mods}.  We
then summarise our microscopic descriptions of star formation and
black hole growth in Sections~\ref{starfeed} and \ref{bhole},
respectively, and discuss aspects of our numerical treatment in
Section~\ref{numerics}.  We use our models to examine the impact of
various forms of feedback on isolated disks in Section~\ref{isostab}
and major mergers in Section~\ref{mmergers}.  Finally, we conclude in
Section~\ref{concs}.

\section{Model Galaxies}
\label{mods}

Each galaxy in our study consists of a dark matter halo, a
rotationally supported disk of gas and stars, and a central (optional)
bulge.  The parameters describing each component are independent, so
that a wide range of morphological types can be specified. The models
are constructed in a manner similar to the approach described by
\citet{Hernquist1993b} and \citet{Springel2000}, but with a number of
refinements.  In particular, we use halo density profiles motivated by
cosmological simulations, and include pressure gradients in setting
the equilibrium structure of the gas. The former consideration is
important for the development of tidal features in interacting systems
\citep{Dubinski1996, Dubinski1999, Mihos1998, Springel1999}, while the
latter improvement is significant when the gas fraction of the disk is
large, or if the gas has a relatively stiff equation of state.

\subsection{Density profiles}
\label{massd}

\subsubsection{Halos}
\label{halomass}

\begin{figure}
\begin{center}
\resizebox{8.cm}{!}{\includegraphics{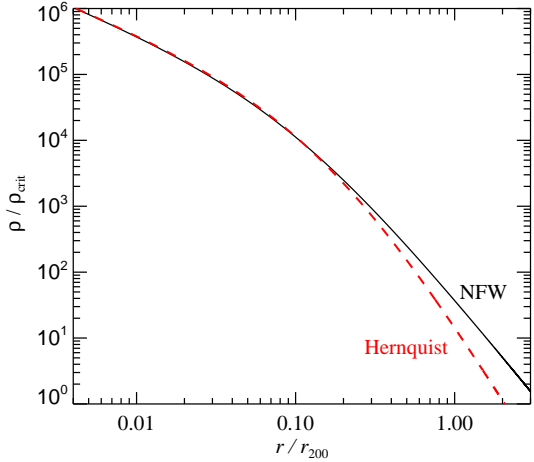}}
\end{center}
\caption{Density profiles of NFW and Hernquist model halos, matched to
each other as described in the text. The halo has concentration
$c=10$. The {\em total} mass of the Hernquist model is equal to he
mass of the NFW profile within the virial radius $r_{200}$.}
\label{fig:halomod}
\end{figure}

In the earlier work of \citet{Hernquist1993b}, halos were described by
truncated isothermal spheres with constant density cores.  However,
N-body simulations such as those of \citet{Dubinski1991} and
\citet{Navarro1996} (hereafter NFW) indicate that cosmological halos
have density profiles that are centrally peaked and drop at large
radii more rapidly with radius than isothermal spheres.

In view of this, we model the dark matter mass distribution with a
\citet{Hernquist1990} profile, i.e.  \be \rho_{\rm dm}(r) =
\frac{M_{\rm dm}}{2\pi}\,\frac{a}{r(r+a)^3}, \ee with cumulative mass
profile $M(<r) = M_{\rm dm}\, r^2/(r+a)^2$. There are several
motivations for this choice. In its inner parts, the shape of this
profile agrees with the NFW fitting formula, but due to its faster
decline in the outer parts, the total mass converges, allowing the
construction of isolated halos without the need for an ad-hoc
truncation. Furthermore, in many situations it is convenient to work
with components having analytical distribution functions, as is the
case for the \citet{Hernquist1990} profile, but not for the NFW model.

To make contact with common descriptions of halos in cosmological
simulations, we associate the Hernquist profile with a corresponding
NFW-halo with the same dark matter mass within $r_{200}$, the radius
at which the mean enclosed dark matter density is 200 times the
critical density. We also require that the inner density profiles are
equal (i.e. $\rho_{\rm dm} = \rho_{\rm NFW}$ for $r\ll r_{200}$),
which implies a relation between $a$ and the scale length $r_s$ of the
NFW profile. The latter is often given in terms of a concentration
index $c$, conventionally defined as $c= r_{200}/r_s$, where $r_s$ is
the scale length of the NFW halo. We then have the relation \be a =
r_s\sqrt{2 [ \ln(1+c) - c/(1+c)]}. \ee For a concentration of $c=10$,
this gives $a\simeq 1.73\, r_s$. The Hernquist profile then contains
75\% of its total mass within $r_{200}$, and 99\% within $1.1\,
r_{200}$.

In Figure~\ref{fig:halomod}, we compare the density profile of an NFW
halo with a Hernquist model of the same concentration.  In the inner
regions, the two mass distributions match closely, while at large
radii the density profiles asymptote to $\rho_{\rm NFW} \propto
r^{-3}$ and $\rho_{\rm Hern} \propto r^{-4}$, respectively.  At
present, observations indicate that either form can give a good
description of the density profiles of actual halos when extended
beyond the virial radius \citep{Rines2000, Rines2002, Rines2003,
Rines2004}.

\subsubsection{Disks and bulges}
\label{diskbulge}

We model disk components of gas and stars with an exponential surface
density profile of scale length $h$; i.e.  \be \Sigma_{\rm gas}(r) =
\frac{M_{\rm gas}}{2\pi h^2} \, \exp(-r/h),
\label{eqGasDens} 
\ee \be \Sigma_{\star}(r) = \frac{M_{\star}}{2\pi h^2} \, \exp(-r/h) ,
\ee so that the total mass in the disk is $M_{\rm d} = (M_{\rm gas} +
M_{\star}) = m_{\rm d} M_{\rm tot}$, where $m_{\rm d}$ is
dimensionless and $M_{\rm tot}$ is the total mass of the galaxy.

The bulge is taken to be spherical, for simplicity.  We also model it
with a Hernquist profile: \be \rho_{\rm b}(r) = \frac{M_{\rm
b}}{2\pi}\,\frac{b}{r(r+b)^3} .  \ee We treat the bulge scale length
$b$ as a free parameter that we parameterise in units of the disk
scale length, and specify the bulge mass as a fraction $m_{\rm b}$ of
the total mass, i.e. $M_{\rm b}= m_{\rm b} M_{\rm tot}$.

We set the disk scale length $h$ by relating it to the angular
momentum of the disk. Following \citet{Mo1998} and
\citet{Springel1999}, we first write the total halo angular momentum
of the halo as \be J = \lambda\, G^{1/2} M_{200}^{3/2} r_{200}^{1/2}
\left(\frac{2}{f_c}\right)^{1/2}, \ee where $\lambda$ is the usual
spin parameter, which we take as a free parameter of our galaxy
models.  The factor \be f_c = \frac{c \left[ 1 - 1/(1+c)^2 -
2\,\ln(1+c)/(1+c)\right]} {2\left[ \ln (1+c) - c/(1+c) \right]^2 } \ee
depends only on the halo concentration index. Of direct relevance for
the structure of the galaxy is the fraction of the angular momentum in
the disk. We typically assume $J_{\rm d} = m_{\rm d} J$, corresponding
to conservation of specific angular momentum of the material that
forms the disk. Assuming that the disk is centrifugally supported,
this then implies a one-to-one relation between $\lambda$ and $h$. In
order to obtain this connection, we compute the disk angular momentum
assuming strict centrifugal support of the disk and negligible disk
thickness compared to its scale-length.  Then we can write \be J_{\rm
d} = M_{\rm d} \int_0^{\infty} V_c(R) \left(\frac{R}{h}\right)^2
\exp\left(-\frac{R}{h}\right)\, {\rm d}R, \ee where the circular
velocity is given by
\begin{eqnarray}
V_c^2(R) &=& \frac{G \left[ M_{\rm dm}(<R) + M_{\rm b}(<R) \right]}{R}
 + \frac{2\,GM_{\rm d} }{h} y^2
\label{eqA}
\\ \nonumber
&&\times\left[I_0(y)  K_0(y) - I_1(y)  K_1(y)\right].
\end{eqnarray}
Here $y=R/(2h)$, and the $I_n$ and $K_n$ are Bessel functions.  Note
that we will later drop the thin-disk approximation for the disk's
potential in favour of an accurate representation of a thick
disk. However, for the conversion of the free parameter $\lambda$ into
a disk scale length $h$, the accuracy of equation (\ref{eqA}) is
sufficient.

We specify the vertical mass distribution of the stars in the disk by
giving it the profile of an isothermal sheet with radially constant
scale length $z_0$. The 3D stellar density in the disk is hence given
by \be \rho_\star(R,z) = \frac{M_{\star}}{4\pi z_0\, h^2}\, {\rm
sech}^2\left(\frac{z}{2\,z_0}\right)\exp\left(-\frac{R}{h}\right).
\ee We treat $z_0$ as a free parameter that effectively determines the
`temperature' of the disk, and set the velocity distribution of the
stars such that this scale-height is self-consistently maintained in
the full 3D potential of the galaxy model.

However, a similar freedom is not available for the gas, because once
cooling and star formation processes are taken into account, we cannot
choose the temperature freely. Rather, in a broad class of models the
gas will stay close to an (effective) equation of state of the form
$P=P(\rho)$, implying a tight relation of gas pressure and gas
density. Examples for such situations include isothermal gas, or the
subresolution multiphase model by \citet{Springel2003a}.  For a given
surface density, the vertical structure of the gas disk then
arises as a result of self-gravity and the pressure given by this
equation of state, leaving no freedom for prescribing a certain
vertical profile. In this situation, the vertical structure has to be
computed self-consistently, as discussed next.

\subsubsection{Vertical structure of the gas disk}

We assume that the vertical structure of the gas disk in our
axisymmetric galaxy models is governed by hydrostatic equilibrium,
i.e.  \be -\frac{1}{\rho_{\rm g}}\frac{\partial P}{\partial z}
-\frac{\partial \Phi}{\partial z} = 0, \ee where $\Phi$ is the total
gravitational potential of all mass components.  This equation can be
rewritten as \be \frac{\partial \rho_{\rm g}}{\partial z} = -
\frac{\rho_{\rm g}^2}{\gamma P}\,\frac{\partial \Phi}{\partial
z}\label{eqB}, \ee where $\gamma = \frac{d\ln P}{d\ln \rho}$ is the
local polytropic index of the equation of state.  Note that for a
given potential $\Phi$, the solution of this equation is determined by
the integral constraint \be \Sigma_{\rm gas}(R,z) = \int \rho_{\rm
g}(R,z)\,{\rm d}z ,
\label{eqC} \ee where $\Sigma_{\rm gas}(r)$ is the surface mass density we
prescribed in equation (\ref{eqGasDens}).  Assuming a given potential for the
moment, we solve equations (\ref{eqB}) and (\ref{eqC}) by integrating the
differential equation for a chosen central density value in the $z=0$
midplane. This chosen starting value is then adjusted until we recover the
desired surface density for the integrated vertical structure of the gas
layer. We carry out this process as a function of radius, using a fine
logarithmic grid of points in the $R-z$ plane to represent the axisymmetric
gas density distribution.

\begin{figure}
\begin{center}
\resizebox{8.0cm}{!}{\includegraphics{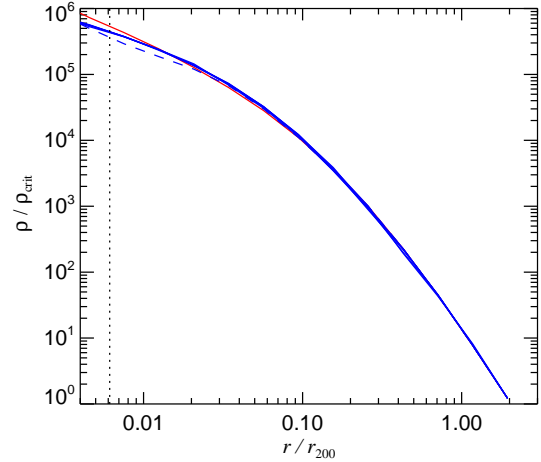}}
\end{center}
\caption{Time evolution of a Hernquist dark matter halo with initial
conditions constructed in the way described in this paper.  The
initially realised input profile is the thin red line.  Because the
approximation of a locally Gaussian velocity distribution is not
exact, the central profile is not in perfect equilibrium in the
beginning. As a result, the density in the centre fluctuates downwards
on a short timescale of order the crossing time, and then relaxes to a
slightly softer central profile.  After time $t=0.08$ (in units of the
dynamical time, $t_{\rm dyn}= r_{200}/v_{200}$, of the halo), the
deviation is close to its maximum (dashed line). Already at $t=0.16$,
however, a stable profile is reached, which then remains essentially
invariant with time, as illustrated by the further output times that
are overplotted (at times $t=0.24$, $0.32$, $0.48$, and $0.64$). The
vertical dotted line marks the scale below which the force law is
softened compared to the Newtonian value.
\label{fig:relaxation}}
\end{figure}

\begin{figure*}
\hbox{
\resizebox{8.5cm}{!}{\includegraphics{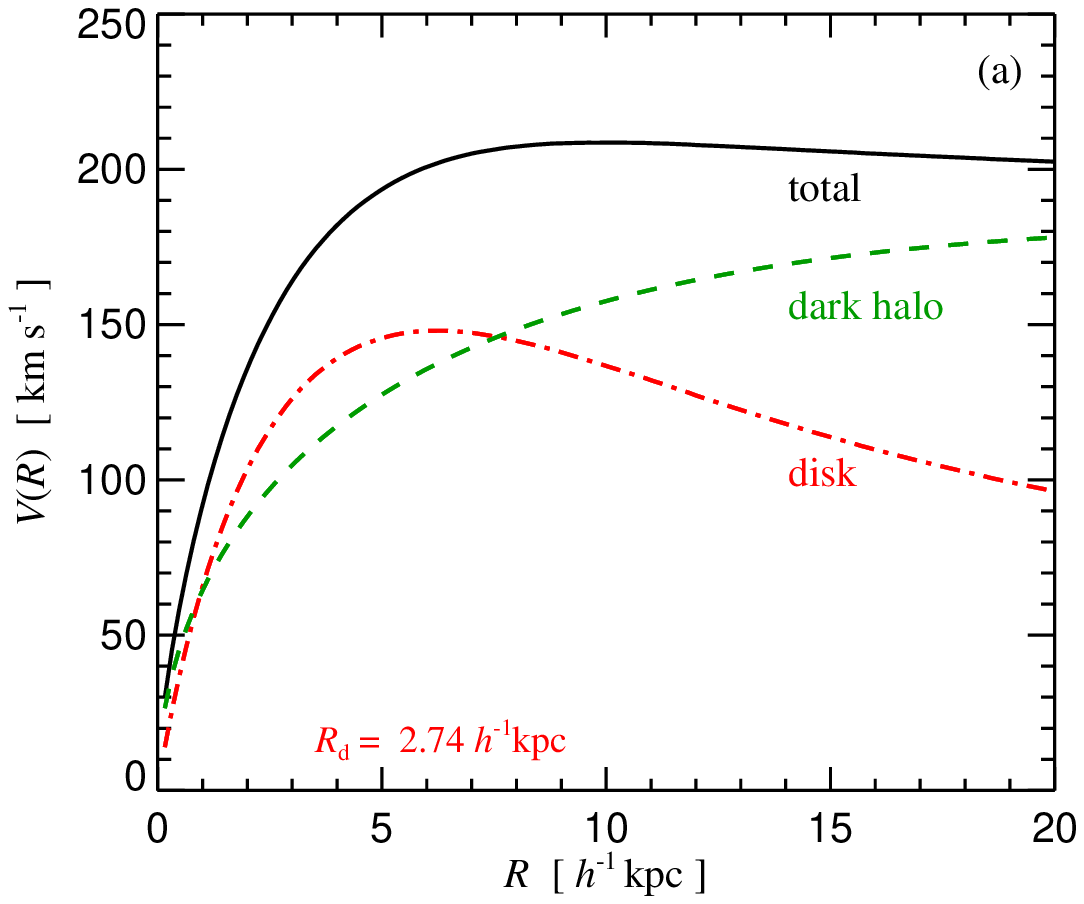}}
\resizebox{8.5cm}{!}{\includegraphics{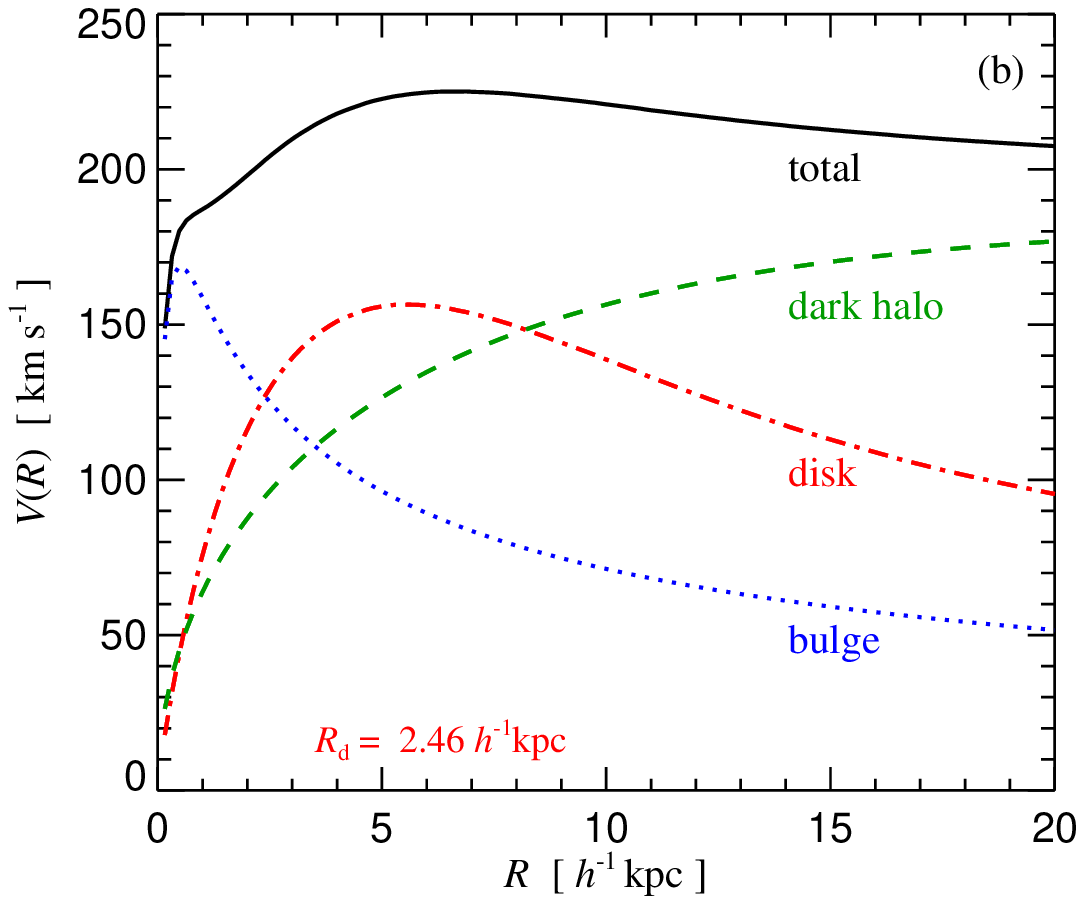}}}
\caption{Rotation curves of model galaxies with the following
parameters: (a) $v_{200} = 160 \kms$, $c=9$, $m_{\rm d} = 0.041$,
$m_{\rm b} = 0$, $f_{\rm gas} = 0.1$, $\lambda = 0.033$, $J_{\rm d} =
0.041$, $z_0 = 0.2$; (b) same as model (a) but with $m_{\rm b} =
0.01367$. The resulting disk scale lengths are $2.74\,h^{-1}{\rm kpc}$
and $2.46\,h^{-1}{\rm kpc}$, respectively. }
\label{fig:galrot}
\end{figure*}

\subsection{Evaluating the potential}

The above invokes the problem of determining the potential and the
resulting gas distribution {\em self-consistently}. We address this
problem by an iterative method. Starting with a guess for an initial
potential, we compute the implied vertical gas structure as described
above, treating the potential as fixed. Then, we recompute the
potential for the updated gas distribution, and repeat the whole
process. In doing this, the potential and the gas distribution
converge quickly, within a few iterations, to a self-consistent
solution.

However, in doing this, we also need a method to accurately compute
the potential and the resulting force field for the radially varying
density stratification of the gas.  Since we want to be able to set up
quite massive gas disks that are in equilibrium right from the start,
we cannot utilise simple thin-disk approximations, but rather need to
compute the full potential accurately. The generality of the gas
distributions used here makes this a non-trivial exercise.

We use a somewhat contrived, yet extremely flexible solution for this
task. We set up a discretised mass distribution that represents the
desired disk components accurately, and then compute the potential
numerically with a hierarchical multipole expansion based on a tree
code.  For a given (current) mass model of the galaxy, we use for this
purpose a suitably distorted grid of particles of typically
$2048\times 64\times 64$ particles per component (the different
numbers refer to radial, azimuthal, and vertical directions,
respectively). While we find that this gives sufficient accuracy for
our purposes, we note that the number of particles can be easily made
much larger to generate a still smoother potential, if needed. Unlike
in a normal N-body code, we are here not interested in evaluating the
forces or the values of the potential for all these particles. We only
use them as markers for the mass, and instead compute the potential at
a fixed set of spatial coordinates. The work required for this
potential computation then scales only with $\log(N)$. The primary
cost of making $N$ large lies therefore in the memory consumed on the
computer, and not in the CPU time.  The potential and force field of
the spherical mass distributions used for the dark matter halo and the
bulge are known analytically, so we simply add them to the result
obtained with the tree for the disk components.

\subsection{Velocity structure}

Once the full density distribution has been determined, we compute
approximations for the velocity structure of the collisionless and
gaseous components.  For the dark matter and stars in the bulge, we
assume that the velocity distribution function only depends on energy
$E$ and the $L_z$-component of the angular momentum. Then, mixed second
order moments of the velocity distribution vanish, $\left<{v_R
v_z}\right> = \left<v_z v_\phi \right>= \left<v_R v_\phi\right> =0$,
as well as the first moments in radial and vertical directions,
$\left<v_R\right> = \left<v_z \right>= 0$.  The velocity distribution
can then be approximated as a triaxial Gaussian with axes aligned with
the axisymmetric coordinate system. The non-vanishing second-moments
can be obtained with the Jeans equations. We have \be
\left<v_z^2\right> = \left<v_R^2\right> = \frac{1}{\rho}\int_z^\infty
\rho(z',R) \frac{\partial \Phi}{\partial z'} \,{\rm d}z', \ee where
$\rho$ is the density of the mass component under consideration.  For
the azimuthal direction we have \be \left<v_\phi^2\right> =
\left<v_R^2\right> + \frac{R}{\rho}\frac{\partial \left(\rho
\left<v_R^2\right>\right)}{\partial R} + v_c^2,
\label{eqJeans}
\ee where \be v_c^2 = R \frac{\partial\Phi}{\partial R}\ee is the
circular velocity.

In the azimuthal direction, there can be a mean streaming component
$\left<v_\phi\right>$, which is not determined by the Jeans equations.
Once it is specified, the dispersion of the Gaussian velocity
distribution in the azimuthal direction is given by \be \sigma_\phi^2
= \left< v_\phi^2 \right> - \left<v_\phi\right>^2. \ee For the stellar
bulge, we set $\left<v_\phi\right>$ to zero, meaning that the bulge is
assumed to have no net rotation. For the dark matter halo, we assume
that it has the same specific angular momentum as the disk material.
For simplicity, we impart this angular momentum by making
$\left<v_\phi\right>$ a fixed fraction of the circular velocity, i.e.
$\left<v_\phi\right> = f_s v_c$ \citep{Springel1999}. Note that we
then have $f_s\ll 1$, i.e.~the net streaming of the resulting dark
halo is quite small.

The velocity structure of the stellar disk can in principle be much
more complicated, and it will in general have a distribution function
that does not only depend on $E$ and $L_z$. For simplicity, we however
continue to approximate the velocity distribution with a triaxial
Gaussian that is aligned with the axisymmetric coordinated vectors.

As Hernquist (1993a) points out, observationally there is good evidence
that $\left<v_R^2\right>$ is proportional to $\left<v_z^2\right>$ for
the stellar disk. We hence assume $\sigma_R^2 = \left<v_R^2\right> =
f_R \left<v_z^2\right>$. Note that larger values of $f_R$ will
increase the Toomre $Q$ parameter, making the disk more stable against
axisymmetric perturbations. In most of our default models, we actually
set $f_R=1$ which corresponds to the approximations made for the dark
halo and the bulge. Note that the Toomre $Q$ also varies in response
to the vertical dispersion $\left<v_z^2\right>$ which sensitively
depends on the assumed disk thickness.

To specify the mean streaming, we employ the epicyclic approximation,
which relates the radial and vertical dispersions in the stellar disk:
\be \sigma_\phi^2 = \frac{\sigma_R^2}{\eta^2}, \ee where \be \eta^2 =
\frac{4}{R} \frac{\partial \Phi}{\partial R} \left(\frac{3}{R}
\frac{\partial \Phi}{\partial R} + \frac{\partial^2 \Phi}{\partial
R^2} \right)^{-1}.  \ee Using equation (\ref{eqJeans}) for
$\left<v_\phi^2\right>$, we then set the streaming velocity for the
stellar disk as \be \left<v_\phi\right> = \left(\left<v_\phi^2\right>
- \frac{\sigma_R^2}{\eta^2}\right)^{1/2}.  \ee This completes the
specification of the velocity structure of the collisionless
components.

For the gas, we deal with a single valued velocity field,
where only the azimuthal streaming velocity has to be specified. The
latter is determined by the radial balance between gravity on one
hand, and centrifugal and pressure support on the other
hand. Specifically, we here have \be v_{\phi,\rm gas}^2 = R \left(
\frac{\partial \Phi}{\partial R} + \frac{1}{\rho_{\rm
g}}\frac{\partial P}{\partial R} \right) .  \ee Note that the gas
temperature at each coordinate is given by the equation of state based
on the self-consistent mass distribution derived above.

We carry out the integrations over the force field needed in the Jeans
equations with the help of a fine logarithmic grid in the
$R$-$z$-plane. This grid is also used to tabulate the final velocity
dispersion fields. The differentiations needed in equation
(\ref{eqJeans}), for example, are approximated by finite-differencing
off this grid.  Once all of the density and velocity distributions
have been computed, we initialise particle coordinates and velocities
by drawing randomly from the respective distributions. Values for the
velocity structure at individual particle coordinates are obtained by
bilinear interpolation of the $R$-$z$ grid.

\citet{Kazantzidis2004} recently examined the accuracy of initialising
spherically symmetric dark matter halos with a Gaussian velocity
distribution with dispersion derived from the Jeans equations. They
found that the innermost parts of halos are not precisely in
equilibrium when this approximation is applied. As a result, the core
relaxes within the first few crossing times to a density profile which
lies slightly below the input profile in the centre, such that the
central cusp becomes softer. Using a more accurate computation of the
velocity distribution function that takes higher order moments into
account, they also demonstrated that this behaviour can in principle
be avoided.

In Fig.~\ref{fig:relaxation}, we show an example of this effect. To
illustrate this in a manner unaffected by two-body relaxation we have
set-up a pure dark matter halo with 4 million particles using our code
for constructing compound galaxies.  When evolved forward in time, the
density profile in the very centre fluctuates below the desired input
value during the first crossing times. Afterwards, the density profile
shows no further secular evolution and stays extremely stable for
times of order the Hubble time. The magnitude of the initial
relaxation effect appears to be consistent with what
\citet{Kazantzidis2004} found. However, we argue that this
perturbation is acceptably small for our purposes. This is because the
effect on the central dark matter profile is really quite moderate,
and occurs in a radial region that is already dominated by baryonic
physics when a disk component is included as well.  In fact, our disk
galaxy models show remarkable little secular relaxation when started
from initial conditions constructed in the above fashion, even for
high gas fractions. We here benefit from the included treatment of gas
pressure forces and the improved potential computation compared to
previous methods \citep{Hernquist1993b,Springel2000}.

\subsection{Galaxy model parameters}

In summary, we specify a disk galaxy model with the following
parameters:
\begin{itemize}
\item The total mass is given in terms of a `virial velocity',
  $v_{200}$. We set $M_{\rm tot} = v_{200}^3 / (10 G H_0)$.
\item The total mass of the disk is given in terms of a dimensionless
  fraction $m_{\rm d}$ of this mass, i.e. $M_{\rm disk} = m_{\rm d}
  M_{\rm tot}$.  Likewise, the mass of the bulge is given in terms of
  a dimensionless fraction $m_{\rm b}$ of the total, i.e. $M_{\rm
  bulge} = m_{\rm b} M_{\rm tot}$. The rest of the total mass is in
  the dark matter halo.
\item A fraction $f_{\rm gas}$ of the disk is assumed to be in gaseous form,
  the rest in stellar form. The bulge is taken to be completely stellar.
\item The stellar disk is assumed to have an exponential profile with
  radially constant vertical scale-height $z_0$. We specify the latter
  in units of the radial disk scale length, and typically use
  $z_0\simeq 0.1-0.2\, h$.
\item Similarly, the bulge scale length is specified as a fraction of
  the radial disk scale length.
\item A value for the spin parameter $\lambda$ of the halo is
  selected. Its value effectively determines the radial scale length
  of the disk.
\end{itemize}

In Figure~\ref{fig:galrot}, we show rotation curves for two galaxy
models that we use in many of the numerical simulations in this
study. The two models have a mass comparable to the Milky Way, $M_{\rm
tot} = 0.98\times 10^{12}\,h^{-1}{\rm M}_\odot$, corresponding to a
virial velocity $v_{200}=160\,{\rm km\, s^{-1}}$. Their disk mass is
specified by $m_{\rm d}=0.041$. The model that includes a bulge (shown
on the right) has a rotation curve very similar to the one used by
\citet{Hernquist1993b}. Note that the model on the left, with its
slowly rising rotation curve due to the absence of a bulge \citep[a
similar model was used, e.g., by][]{Mihos1994a}, is more susceptible
to tidal perturbations.

\begin{figure}
\begin{center}
\resizebox{8.0cm}{!}{\includegraphics{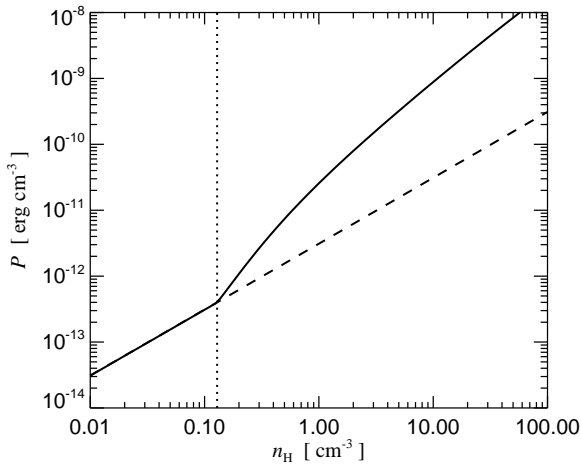}}
\end{center}
\caption{Effective equation of state for the star-forming gas (solid
curve).  We show the effective pressure measured in cgs units as a
function of the gas density in hydrogen atoms per cubic centimetre.
The vertical dotted line delineates the transition from an isothermal
gas (dashed) at temperature $10^4\,{\rm K}$ to the regime governed by
an effective pressure in our multiphase model.  For the particular
parameters chosen in this example, the transition density lies at
$0.128\, {\rm cm^{-3}}$.}
\label{fig:eosfig}
\end{figure}

\section{Star Formation and Stellar Feedback}
\label{starfeed}

Current state of the art simulations lack the dynamic range needed to
follow the detailed structure of star-forming gas on scales
characterising entire galaxies.  For example, in simulations like
those presented in the following sections of this paper, tens or
hundreds of
thousands of particles are used to characterise the gas in individual
disk galaxies, so that each particle represents roughly $\simgt 10^5\,
{\rm M}_\odot$ of interstellar material for typical disk gas
fractions. However, in order to resolve bound structures in SPH, an
object must contain at least a number of particles comparable to the
number averaged over when forming smoothed quantities.  Here, we set
this parameter typically to be $\simeq 64$ neighbouring particles, so
that the simulations cannot resolve bound objects with masses smaller
than about $\sim 6\times 10^6 {\rm M}_\odot$.  Simulations with total
particle numbers 100 times larger than those presented here are
feasible, but even then we would not able to resolve self-gravitating
structures with masses $\simlt 10^5\, {\rm M}_\odot$.  Furthermore,
such calculations would be limited to a spatial resolution of order
$\sim 20$ pc.  These mass and spatial scales are barely sufficient to
resolve individual giant molecular clouds and are inadequate to
describe the details of the star-forming gas contained within them.

More seriously, there is at present no complete theory for star formation, so
the important physics is not well understood.  Even if the simulations could
include the entire dynamic range needed to resolve the essentials of star
formation on galactic scales, the description of this process would therefore
still have to be parameterised using bulk averaged quantities.  For these
reasons, we are motivated to adopt an approach in which we discard detailed
information about star-forming gas in favour of a procedure that captures
generic consequences of star formation and associated feedback effects on
resolved scales.  In this manner, we do not attempt to characterise all
aspects of the physics but instead seek to identify general trends.

As an illustration of our procedure, we employ the subresolution
multiphase model for star-forming gas developed by \citet[][hereafter
SH03]{Springel2003a}.  In this picture, the ISM is pictured to consist
of cold clouds where stars form, embedded in a hot, pressure-confining
phase.  The gas is assumed to be thermally unstable to the onset of a
two-phase medium at densities above a threshold $\rho_{\rm th}$.  The
fraction of mass in the two phases at higher densities is set by star
formation and feedback, evaporation of the cold clouds through thermal
conduction, and the growth of clouds owing to radiative cooling.  The
rates at which the mass in the two phases evolve are given by
eqns.~(5) and (6) in SH03.  The thermal budget of the cold clouds and
hot gas is determined by the same effects and evolves according to
eqns.~(8) and (9) in SH03.

Our coarse-grained representation of this model reduces largely to two
principle ingredients: (1) a star formation prescription or ``law'',
and (2) an effective equation of state (EOS) for the ISM.  For the
former, we adopt a rate motivated by observations
\citep{Kennicutt1989, Kennicutt1998}: \be {{{\rm d}\rho _*}\over{{\rm
d}t}} \, = \, (1 \, - \, \beta) \, {{\rho_c}\over t_*} , \ee where
${\rm d}\rho _*/{\rm d}t$ is the star formation rate, $\beta$ depends
on the initial mass function (IMF) and is the mass fraction of stars
that die immediately as supernovae, $\rho_c$ is the density of cold
clouds, and $t_*$ is a characteristic timescale.  For a Salpeter type
IMF we have $\beta \approx 0.1$.  In our multiphase picture, $\rho_c
\approx \rho$, where $\rho$ is the total gas density (see Fig.~1 in
SH03).  A good match to observations is obtained if $t_*$ is assumed
to be proportional to the local dynamical time \be t_*(\rho) \, = \,
t_*^0 \, \left ( {{\rho}\over{\rho_{\rm th}}} \right ) ^ {-1/2} , \ee
where $t_*^0$ is a constant parameter.

The effective equation of state (EOS) of the gas in this model is
given by \be P_{\rm eff} \, = \, (\gamma - 1) \, (\rho_h u_h \, + \,
\rho_c u_c), \ee where $\rho_h$ and $u_h$ are the density and specific
thermal energy of the hot phase, with similar quantities defined for
the cold clouds, while $\gamma=5/3$ is the adiabatic index of the gas.
In detail, the EOS depends on the values of the parameters in our
model.

As described by SH03, the parameter values are constrained by the
nature of the cooling curve for the gas and by requiring that the EOS
be a continuous function of density.  SH03 also showed that if star
formation is rapid compared with adiabatic heating or cooling owing to
the motion of the gas, then the multiphase picture leads to a cycle of
self-regulated star formation where, in equilibrium, the growth of
cold clouds is balanced by their evaporation from supernova feedback.
Under these conditions and for a given IMF, the EOS is determined by
three parameters: the constant appearing in the star formation law,
$t_*^0$, a normalisation of the cloud evaporation rate, $A_0$, and a
supernova ``temperature'', $T_{\rm SN}$, that reflects the heating
rate from a population of supernova for the adopted IMF.  The
requirement that the gas be thermally unstable at densities above
$\rho_{\rm th}$ fixes the ratio $T_{\rm SN}/A_0 \approx 10^5$ while in
the self-regulated regime, the condition that the EOS be continuous
implies that $\rho_{\rm th}$ depends mainly on the ratios $T_{\rm
SN}/A_0$ and $T_{\rm SN}/t_*^0$ (see eqn.~23 of SH03).

In Figure~\ref{fig:eosfig}, we show an example of the form of our
effective EOS for the parameter choices $T_{\rm SN} = 10^8$, $A_0 =
1000$, and $t_*^0 = 2.1$ Gyrs.  For an isolated disk galaxy, these
values reproduce the observed correlation between star formation and
gas density (see Figs. 2-3 of SH03).  For densities higher than
$\rho_{\rm th}$, the EOS shown in Figure~\ref{fig:eosfig} is fitted to
an accuracy of $1\%$ by \citep{Robertson2004}
\begin{eqnarray}
\log P_{\rm eff} \,& =& \, 0.050 \, (\log n_{\rm H})^3 \, - \, 
0.246 \, (\log n_{\rm H})^2 + \, \\\nonumber
&&+ \, 1.749 \, \log n_{\rm H} \ - \, 10.6 \, .
\label{eq:eosfit}
\end{eqnarray}
The important features of our EOS can be seen from
Figure~\ref{fig:eosfig}.  If the gas were isothermal at all densities,
the effective pressure, $P_{\rm eff}$, would increase linearly with
density, as indicated by the dashed line for densities higher than
$\rho_{\rm th}$.  However, when we account for supernova feedback in
our multiphase model, the EOS in star-forming gas is stiffer (solid
curve in Figure~\ref{fig:eosfig}).  Suppose a region at densities
lower than $\rho_{\rm th}$ is compressed.  If the gas were isothermal,
the effective pressure would increase proportional to the density.
However, in our multiphase model, once $\rho > \rho_{\rm th}$, the gas
is thermally unstable and stars can form.  As the density rises
further, the star formation rate increases even faster, as does the
rate of energy injection into the gas from supernova.  If this
feedback energy is retained by the gas, its effective EOS of state
will be stiffer than isothermal.

The concept of an ``effective equation of state'' thus incorporates
the impact of local feedback in regulating the thermodynamic
properties of the ISM. Using this concept, we can describe the
dynamics of star-forming gas on galactic scales and account for the
consequences of stellar feedback on large scales.  For example, the
dynamical stability of a pure gas disk to axisymmetric perturbations
is determined by the condition \citep{Toomre1964} \be Q \, \equiv \,
{{c_s \kappa}\over{\pi G \Sigma}} \,\, > \, 1, \ee where $c_s$ is the
sound speed, $\kappa$ is the epicyclic frequency, and $\Sigma$ is the
gas surface density.  This relation represents a competition between
the stabilising influences of pressure ($c_s$) and angular momentum
($\kappa$) against the destabilising effect of gravity ($\Sigma$).  If
we consider a small region in the disk, the appropriate EOS with which
to compute $c_s$ is not one that captures the detailed structure of
the ISM, but an effective EOS on the scale of the instability.  For
parameters typical of galactic disks, the fastest growing mode has a
wavelength that is comparable to the size of the disk
\citep{Binney1987}, and so stability is determined by the response of
the ISM to compressions on this scale.

In the absence of a complete theory for the ISM, the choices for the star
formation law and effective equation of state are not unique.  For example,
\citet{Barnes2004} has recently proposed a star formation prescription that is
based on the rate at which gas generates entropy in shocks.  His numerical
tests indicate that this formalism may yield a better match to observations of
interacting galaxies than a purely density-dependent star formation
law.  

In our approach, we will consider the macroscopic star formation law
and effective equation of state as elements of our description that
can in principle be varied independently to isolate the large-scale
physics that regulates star formation in colliding and merging
galaxies.  In particular, we will also consider simulations where we
``soften'' the EOS by linearly interpolating between that for the
standard implementation of our multiphase model and an isothermal EOS.
We will denote this softening by a parameter $q_{\rm EOS}$ so that 
$q_{\rm EOS}=1$ will
correspond to e.g.~the ``stiff'' EOS shown by the
solid curve in Figure~\ref{fig:eosfig},
and $q_{\rm EOS}=0$ will correspond to the ``soft,'' isothermal EOS,
indicated by the
dashed curve in Figure~\ref{fig:eosfig}.

While we do not expect that the effective EOS employed here is a
precise description of real star-forming gas, the approach that we
adopt is flexible and can be applied to any well-specified theory for
the ISM.  In particular, \citet{Springel2003b} have shown that our
coarse-graining procedure leads to a numerically converged estimate
for the cosmic star formation history of the Universe
\citep{Hernquist2003} that agrees well with low redshift observations.

The hybrid multiphase model of SH03 also includes a phenomenological
representation of galactic winds.  In this picture, some of the energy
available from supernova feedback is tapped to drive an outflow if the
star formation rate exceeds a critical threshold. For simplicity, we
do not include this additional mode of feedback in the present study.

\section{Black Hole Accretion and AGN Feedback}
\label{bhole}

It is now widely believed that black hole growth and associated
feedback energy from this process may be important for a variety of
phenomena related to the evolution of ordinary galaxies as well as
unusual behaviour in quasars, starbursts, and AGN. It has been
suggested that the $M_{\rm BH} - \sigma$ relation may arise if strong
outflows are produced in response to a major phase of black hole
accretion, which via their interaction with the surrounding gas would
inhibit any further accretion and hence black hole growth
\citep{Silk1998, Fabian1999, King2003, Wyithe2003}.  Indeed, X-ray
observations of a number of quasars (mostly broad absorption line
systems) reveal significant absorption, implying large outflows with a
kinetic power corresponding to a significant fraction of the AGN
bolometric luminosity \citep{Chartas2003, Crenshaw2003, Pounds2003}.
In the case of radio-loud QSO, there is also evidence that up to half
of the total power is injected in the form of jets
\citep{Rawlings1991, Tavecchio2000}.  Inevitably, such outflows must
have a strong impact on the host galaxies.  Models of interacting and
merging galaxies should, therefore, account for feedback from both
star formation and accretion onto central, supermassive black holes.
Note that black hole (BH) 
accretion is expected to be responsible for the majority
of the BH mass growth and to provide sufficient energy supply for
driving associated outflows.

As with star formation, current numerical simulations cannot resolve
the properties of the accretion flow around nuclear black holes on
galactic scales. For example, consider a black hole of mass $M_{\rm
BH}$ accreting spherically from a stationary, uniform distribution of
gas whose sound speed at infinity is $c_\infty$. The gravitational
radius of influence of the black hole is then \citep{Bondi1952} \be
r_{\rm B} \, = \, {{G M_{\rm BH}}\over {c_\infty^2}} \, .  \ee For a
black hole of mass $M_{\rm BH} = 10^7 \, {\rm M}_\odot $ in a gas with
a sound speed $c_\infty = 30 \, {\rm km/sec}$, this is numerically \be
r_{\rm B} = 50 \,{\rm pc} \, \left(\frac{M_{\rm BH}}{10^7\,{\rm
M}_\odot}\right) \, \left(\frac{c_\infty}{30 \, {\rm
km/sec}}\right)^{-2}.  \ee The Schwarzschild radius of a black hole of
this mass is \be r_{\rm s} \, \equiv \, {{2 G M_{\rm BH}}\over{c^2}}
\, = \, 10^{-6} \,{\rm pc} \left(\frac{M_{\rm BH}}{10^7\,{\rm
M}_\odot}\right).  \ee Our understanding of the nature of accretion
onto supermassive black holes is sufficiently poor that it is not
clear what range in spatial scales would be required to obtain an
accurate description of the impact of black hole growth and feedback
on galactic scales.  However, given the relatively poor resolution
that can be achieved in simulations like those presented here, it is
clear that they cannot represent the full complexities of this process
in any detail.  By analogy with out treatment of star formation and
supernova feedback, we are hence led to adopt a coarse-graining
procedure.

In what follows, we use an effective sub-resolution model to
characterise the growth of supermassive black holes in galactic nuclei
and the consequences of feedback from accretion on surrounding gas.
Technically, we represent black holes by collisionless particles that
can grow in mass by accreting gas from their environments. A fraction
of the radiative energy released by the accreted material will be
assumed to couple thermally to nearby gas and influence its motion and
thermodynamic state.

Like our procedure for coarse-graining the ISM, our method is flexible
and can be applied to any model for black hole accretion.  As a
starting point, we relate the (unresolved) accretion onto the black
hole to the large scale (resolved) gas distribution using a
Bondi-Hoyle-Lyttleton parameterisation \citep{Bondi1952,
BondiHoyle1944, Hoyle1939}.  In this description, the accretion rate
onto the black hole is given by \be \dot{M}_{\rm B} \, = \, {{4\pi \,
\alpha \, G^2 M_{\rm BH}^2 \, \rho} \over {(c_s^2 + v^2)^{3/2}}} \, ,
\ee where $\rho$ and $c_s$ are the density and sound speed of the gas,
respectively, $\alpha$ is a dimensionless parameter, and $v$ is the
velocity of the black hole relative to the gas.  We will also assume
that the accretion is limited to the Eddington rate \be \dot{M}_{\rm
Edd} \, \equiv \, {{4\pi \, G \, M_{\rm BH} \, m_{\rm p}} \over
{\epsilon_{\rm r} \, \sigma_{\rm T} \, c}} \, , \ee where $m_{\rm p}$
is the proton mass, $\sigma_{\rm T}$ is the Thomson cross-section, and
$\epsilon_{\rm r}$ is the radiative efficiency. The latter is related
to the radiated luminosity, $L_{\rm r}$ and accretion rate, $\dot
{M}_{\rm BH}$, by \be \epsilon_{\rm r} \, = \, {{L_{\rm r}}\over{\dot
{M}_{\rm BH} \, c^2}} \, , \ee i.e.~it simply gives the mass to energy
conversion efficiency set by the amount of energy that can be
extracted from the innermost stable orbit of an accretion disk around
a black hole.  For the rest of this study, we adopt a fixed value of
$\epsilon_{\rm r} =0.1$, which is the mean value for radiatively
efficient \citet{Shakura1973} accretion onto a Schwarzschild
black hole. We ignore the possibility of radiatively inefficient
accretion phases. The accretion rate is then \be \dot{M}_{\rm BH} \, =
\, {\rm min} (\dot{M}_{\rm Edd}\, , \, \dot{M}_{\rm B}) \, .  \ee

We will assume that some fraction $\epsilon_{\rm f}$ of the radiated
luminosity $L_{\rm r}$ can couple thermally (and isotropically) to
surrounding gas in the form of feedback energy, viz. \be \dot{E}_{\rm
feed} \, = \, \epsilon_{\rm f} \, L_{\rm r} \,\, = \, \epsilon_{\rm f}
\, \epsilon_{\rm r} \, \dot{M}_{\rm BH} \, c^2 \, .  \ee
Characteristically we take $\epsilon_{\rm f} \sim 0.05$, so that $\sim
0.5 \%$ of the accreted rest mass energy is available as feedback.
This value fixes the normalisation of the $M_{\rm BH} - \sigma$
relation, and brings it into agreement with current observations
\citep{DiMatteo2004}.

\section{Numerical approach}

\label{numerics}

\begin{figure}
\begin{center}
\resizebox{8.0cm}{!}{\includegraphics{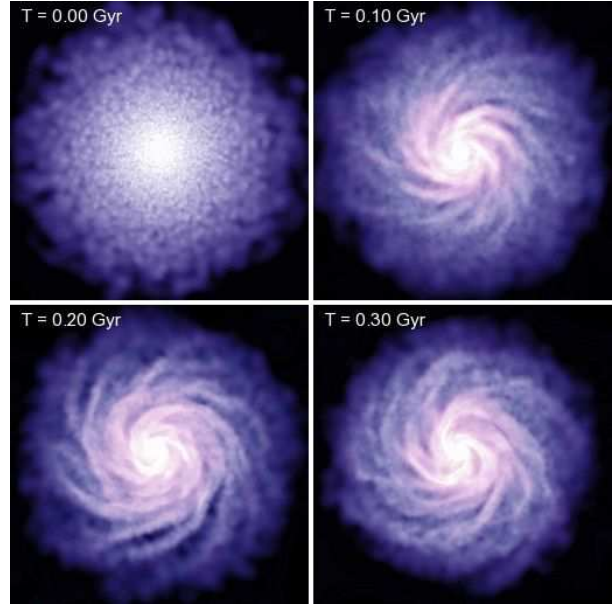}}
\end{center}
\caption{Evolution of the gaseous disk in an isolated galaxy model
  with a bulge component.  A fraction $f_{\rm gas} = 0.1$ of the disk
  mass is in gas.  The remaining $90\%$ is in old stars.  The panels
  show a face-on projection of the gas in the disk, and measure
  $30\,h^{-1}{\rm kpc}$ on a side. The colour-coding indicates both
  gas density (in terms of intensity) and local gas temperature (in
  terms of colour hue). Time in Gyrs is indicated in the upper left
  corner of each frame.}
\label{fig:isodisk}
\end{figure}

Nothing in our formulation of supernova and black hole feedback is
tied to a particular numerical scheme for solving the dynamical
equations for the gas and collisionless matter.  Our sub-resolution
models and coarse-graining procedure could be incorporated into either
particle- or grid-based methods.  In our simulations, we employ an
N-body algorithm for the collisionless material which, in our case,
includes dark matter, stars, and black holes.  For the hydrodynamics,
we employ a smoothed particle hydrodynamics (SPH) code which
represents fluids elements by particles \citep{Lucy1977, Gingold1977,
Monaghan1992}.  In SPH, fluid properties at a given point are
estimated by local kernel-averaging over neighbouring particles, and
smoothed versions of the equations of hydrodynamics are solved for the
evolution of the fluid.

The particular code we use is an improved and updated version of
{\small GADGET} \citep{Springel2001}. The code implements a TreeSPH
algorithm \citep{Hernquist1989} where gravitational forces are
computed with a hierarchical tree method \citep{BarnesHut1986} and SPH
is used for the hydrodynamics.  In the new version {\small GADGET2}
employed here, the hydrodynamical equations are solved using a fully
conservative technique \citep{Springel2002}, which maintains strict
entropy and energy conservation even when smoothing lengths vary
adaptively \citep{Hernquist1993c}.

Besides gravity and hydrodynamics, the code follows radiative cooling
processes of an optically thin primordial plasma of helium and
hydrogen in the presence of an UV background \citep{Katz1996}. Star
formation and the dynamics of the highly overdense gas are treated
with the multiphase model described earlier.  Independent star
particles are stochastically spawned out of the gas phase
\citep{Springel2003a}, thereby avoiding artificial coupling between
gaseous and collisionless matter.

Black holes are represented by collisionless ``sink'' particles which
only feel gravitational forces. For these particles, we compute
estimates for the gas density in their local environments, in the same
fashion as it is done for normal SPH particles. Similarly, we also
determine the average gas temperature in the local SPH smoothing
environment around black hole particles, as well as the gas bulk
velocity relative to the black holes.  These quantities are then used
to estimate the black hole accretion rates, based on the equations
specified in the previous section.

To implement the actual accretion, we follow a similar stochastic
approach as it is applied for regular star formation.  To this end, we
compute for each gas particle $j$ around a black hole a probability
\be p_j = \frac{w_j \dot M_{\rm BH}\Delta t }{\rho} \ee for being
absorbed by the BH. Here $\dot M_{\rm BH}$ is the BH accretion rate,
$\Delta t$ is the timestep, $\rho$ is the gas density estimated at the
position of the black hole, and $w_{j}$ is the kernel weight of the
gas particle relative to the BH.  We then draw random numbers $x_j$
uniformly distributed in the interval $[0,1[$ and compare them with
the $p_j$. For $x_j<p_j$, the gas particle is absorbed by the black
hole, including its momentum. On average, this procedure ensures that
the BH particle accretes the right amount of gas consistent with the
estimated accretion rate $\dot M_{\rm BH}$.

However, because the accretion rate depends sensitively on $M_{\rm
BH}$, this procedure would be quite noisy under conditions of poor
resolution, where accreting a single gas particle can change the BH
particle mass by a substantial factor. We circumvent this problem by
giving the sink particle an internal degree of freedom that describes
the BH mass in a smooth fashion. The value of this variable represents
the BH mass for the case of ideal resolution where the gas mass is not
discretised.  The real dynamical mass of the sink particle tracks this
internal mass closely, albeit with stochastic fluctuations around
it. For this reason, we use the internal mass for computing the
accretion rates, while the growth of the hole is followed both in
terms of the internal mass and the actual dynamical mass. The latter
increases in discrete steps when whole gas particles are absorbed, but
follows the internal variable in the mean, with fluctuations that
become progressively smaller for better resolution. Using this
procedure we can reliably follow the growth of black holes in terms of
their internal mass variable even in halos with just a few hundred
particles, which is particularly important for cosmological
simulations, where the earliest generations of galaxies are typically
not very well resolved. Note that we conserve momentum when gas
particles are absorbed by BH sink particles. If the BH is moving
relative to the gas and has a high accretion rate, this can
effectively act like a friction force which reduces the relative
motion.

Together with the accretion, we compute the rate of feedback
associated with the black hole growth. This energy is added
kernel-weighted to the thermal reservoir of the gas in the local
environment around the black hole.  Note that if the local cooling
rate of the gas is not high enough to radiate away all of this energy
on a short timescale, an increase of the gas sound speed occurs which
can then throttle the accretion in the Bondi-dominated regime.

In the final stages of a galaxy merger, the cores of the galaxies
coalesce to form a single dark matter halo, and a single stellar
system. Presumably, this also means that a central binary system of
two supermassive black holes is formed, which may subsequently harden
and eventually lead to physical collision and merger of the black
holes themselves. However, it is a controversial question how long it
would take to harden the black hole binary by stellar-dynamical
\citep{Makino2004} or hydrodynamical processes \citep{Escala2004}
until finally gravitational wave emission becomes important, leading
to quick orbital decay. If black hole merger processes were
inefficient, it should be a common situation that a third supermassive
black hole is brought in by the next galaxy merger. This black hole
could then interact with the binary and lead to sling-shot ejection of
one of the holes, with the remaining pair being ejected in the
opposite direction. As this may seriously impair the growth of black
holes to the large masses that are observed, the existence of
supermassive black holes may be taken as circumstantial evidence that
BH binaries probably do merge.  We therefore assume that two black
hole particles merge instantly if they come within each others's
smoothing lengths, and their relative velocities are at the same time
smaller than the local sound speed.  Note that as with the accretion
itself, we lack the dynamic range in simulations of whole galaxies to
study the hardening process of black hole binaries directly.

\begin{figure*}
\begin{center}
\resizebox{17.5cm}{!}{\includegraphics{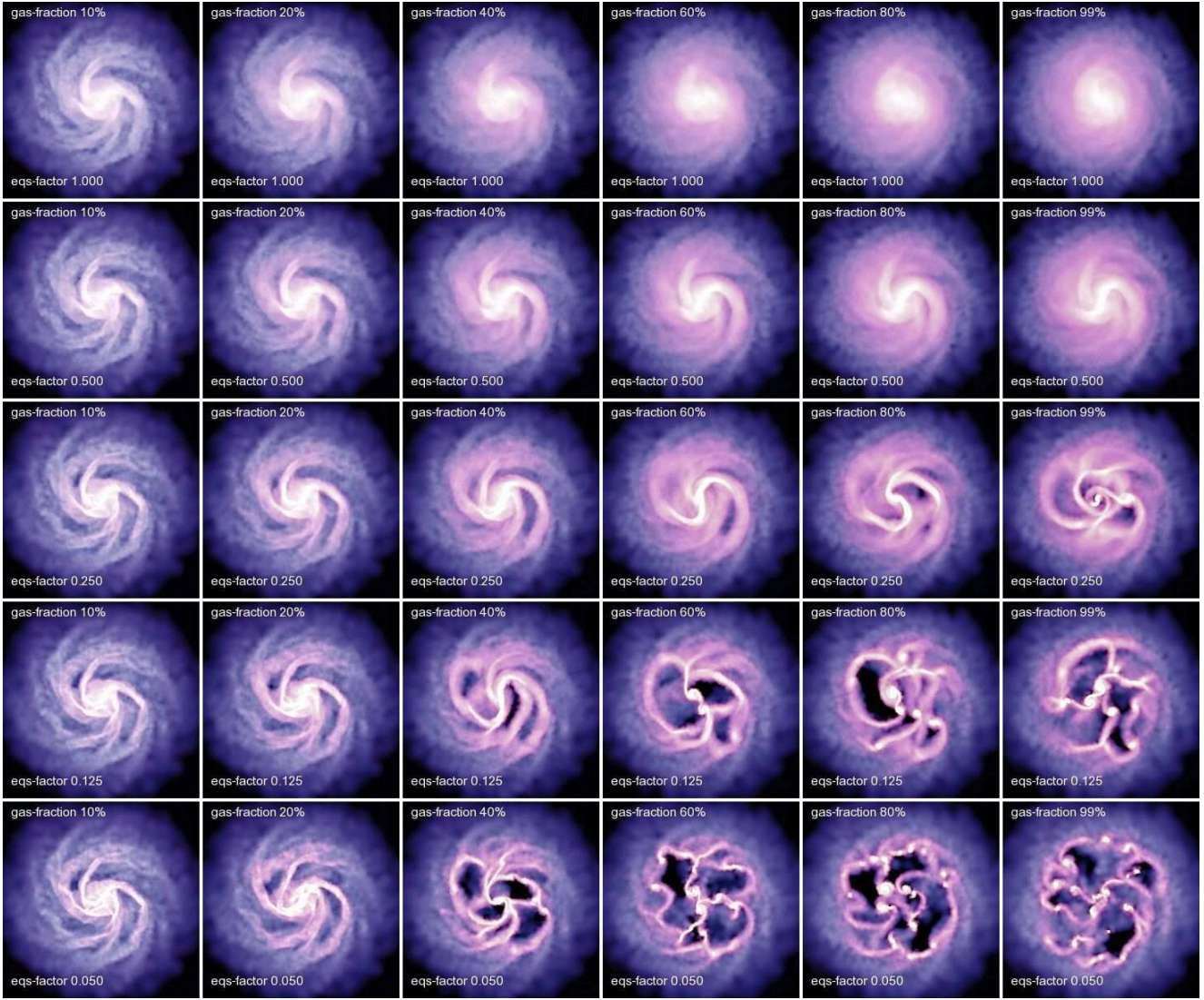}}
\end{center}
\caption{Stability analysis of disk galaxies as a function of gas
  fraction in the disk, $f_{\rm gas}$, and stiffness of the effective
  EOS. Each panel shows the face-on distribution of gas obtained at
  time $t=0.3$ Gyrs after the start of the simulations, roughly
  corresponding to one rotation period at three exponential
  scale-lengths.  The gas fraction is constant along columns, while
  each row shows the results for a fixed EOS. In all cases, the
  initial galaxy models had the same structure, except for a different
  partitioning of the mass in the disk between a gaseous and a stellar
  component. The galaxies in these examples did not include stellar
  bulges, but the results for galaxy models with bulges are
  qualitatively very similar.}
\label{fig:isomontage}
\end{figure*}

\section{Isolated Disk Galaxies}
\label{isostab}

We have run a series of simulations of isolated galaxies using the
methods described in Section~\ref{mods} to initialise them.  In the
following, we examine their stability using models with and without
bulges, having the rotation curves shown in Figure~\ref{fig:galrot}.
The structure of these models was motivated by properties of galaxies
like the Milky Way and to enable us to compare our results with
earlier work, such as \citet{Hernquist1995} and \citet{Mihos1996}.
Here, we focus on the effect of varying the stiffness of the EOS and
the gas fraction in the disk.  A significant advantage of our
formalism is that we are now able to construct stable equilibrium
models with a larger gas fraction than possible in previous studies.

\subsection{Stability of models without black holes}
\label{modsnobh}

In order to facilitate comparison between different cases, in this
section we fix the parameters associated with our multiphase model and
vary only the disk gas fraction and the stiffness of the effective
EOS.  For definiteness, we take $t_*^0 = 8.4\,{\rm Gyrs}$, $A_0 =
4000$, and $T_{\rm SN} = 4\times 10^8$.  This choice for $t_*^0$ is a
factor of four larger than that employed by SH03 to match the
Kennicutt Law.  However, we find that for our isolated galaxies,
$t_*^0 = 2.1\,{\rm Gyrs}$ yields global star formation rates $\simgt
4\, {\rm M}_\odot/{\rm yr}$ for gas fractions of $10\%$ and structural
properties similar to the Milky Way.  In our Galaxy, the inferred
average star formation rate is only $\sim 1 \,{\rm M}_\odot/{\rm yr}$.
Increasing $t_*^0$ by a factor of four gives therefore better
agreement with the long gas consumption timescale inferred for the
Galaxy.  This also simplifies a comparison with \citet{Hernquist1995}
and \citet{Mihos1996}, who chose parameters in their star formation
prescription to give global star formation rates $\sim 1\, {\rm
M}_\odot/{\rm yr}$ for galaxies like the Milky Way with $10\%$ of
their disk mass in gas.

Once $t_*^0$ is fixed, we select $A_0$ and $T_{\rm SN}$ as above so
that the critical density for the gas to be thermally unstable,
$\rho_{\rm th}$, implies a critical projected gas surface density for
star formation to occur that is similar to observations
\citep{Kennicutt1989, Kennicutt1998}.  Since $\rho_{\rm th}$ in our
formalism depends mainly on the ratios $T_{\rm SN}/A_0$ and $T_{\rm
SN}/t_*^0$, a factor of four lengthening of $t_*^0$ relative to SH03
requires a factor of four increase in $T_{\rm SN}$ and $A_0$ for a
critical gas surface density $\sim 10\, {\rm M}_\odot / {\rm pc}^{-2}$
(see Figs. 2-3 of SH03).  As we indicated earlier, the effective EOS
depends primarily on ratios between these three parameters.  Scaling
$t_*^0$, $T_{\rm SN}$, and $A_0$ by the same factor implies that the
EOS for our isolated disks is the same as the ``stiff'' case in
Figure~\ref{fig:eosfig}, only the gas consumption timescale is
larger. In the stability analysis shown below, we hold these
parameters fixed, so that the threshold density for star formation is
the same, but we artificially vary the stiffness of the EOS by
linearly interpolating between the stiff and isothermal limits in
Figure~\ref{fig:eosfig}.  EOS softenings of $q_{\rm EOS}=1$ 
and $q_{\rm EOS}=0$ refer to
the stiff and isothermal cases, respectively.  We also vary the
fraction $f_{\rm gas}$ of the disk mass in gas, but keep the total
disk mass constant.

\begin{figure}
\begin{center}
\resizebox{8.0cm}{!}{\includegraphics{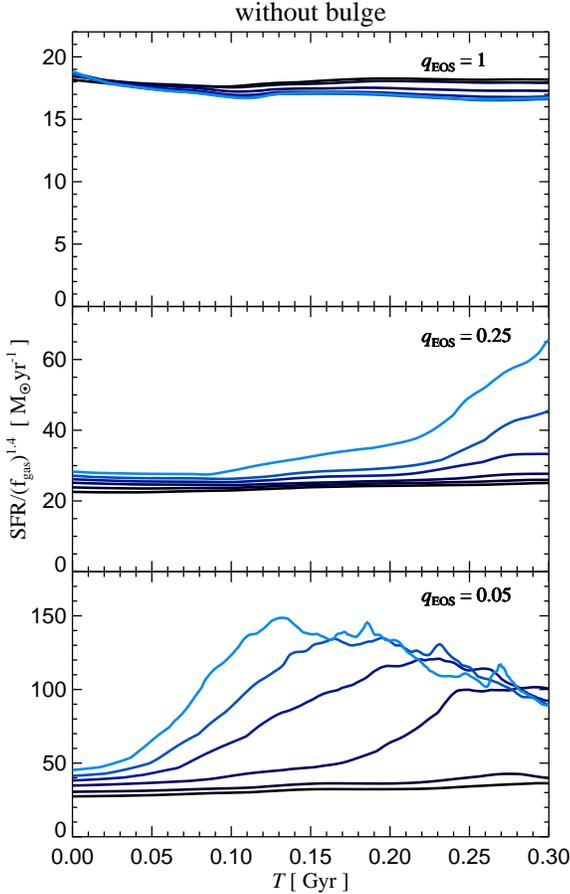}}
\end{center}
\caption{Evolution of the global star formation rate in galaxies
without bulges, as a function of the gas fraction and EOS used.  The
star formation rate in solar masses per year is scaled by the disk gas
fraction raised to the 1.4 power (as in the Kennicutt Law).  Each
panel shows the evolution for the six values of $f_{\rm gas}$ shown in
Figure~\ref{fig:isomontage}, with lighter colours corresponding to
larger gas fractions.  From top to bottom, the panels differ in the
employed EOS for the star forming gas. Results are given for the stiff
EOS of the multiphase model ($q_{\rm EOS}=1$), 
an intermediate case between this
model and an isothermal EOS ($q_{\rm EOS}=0.25$), 
and a `soft' model with a
nearly isothermal EOS ($q_{\rm EOS}=0.05$).  }
\label{fig:isosfrnob}
\end{figure}

\begin{figure}
\begin{center}
\resizebox{8.0cm}{!}{\includegraphics{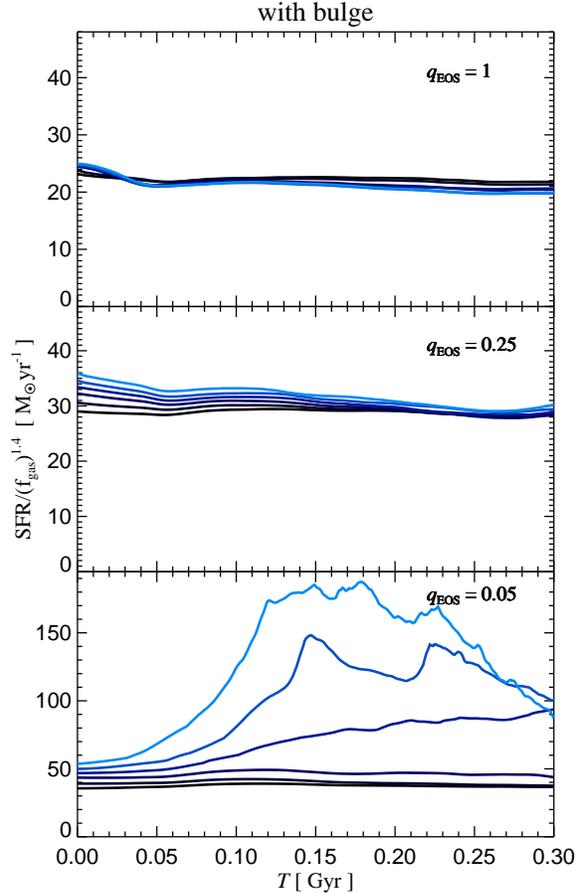}}
\end{center}
\caption{Same as in Figure~\ref{fig:isosfrnob}, but for galaxy models
  that include bulges.}
\label{fig:isosfrb}
\end{figure}

As a typical example, Figure~\ref{fig:isodisk} shows the time
evolution of the projected gas density in an isolated galaxy model
with a bulge.  The gas fraction is $f_{\rm gas} = 0.1$ and the EOS
softening parameter is $q_{\rm EOS}=0.5$.  Initially, the particles randomly
sample the exponential surface density distribution, so there are no
coherent structures present in the disk. However, a steady spiral
pattern develops within a fraction of a rotation period owing to the
action of swing amplification \citep{Toomre1981}.  This behaviour is
similar to that reported in previous numerical studies of disk
structure. For the parameter choices in this example, the disk is
stable for many rotation periods and does not evolve strongly.  The
gas is converted into stars at a relatively low rate according to our
star formation prescription but, otherwise, the disk, bulge, and halo
do not evolve.

However, strong, unstable evolution is possible if the gas fraction is
large and the EOS is too soft.  In Figure~\ref{fig:isomontage}, we
compare the distribution of gas in model galaxies without bulges after
approximately one rotation period of evolution.  The gas fraction is
increased from $f_{\rm gas} = 0.1$ to $f_{\rm gas} = 0.99$ along rows
(left to right) while the EOS is softened from $q_{\rm EOS}=1$ (top) 
to $q_{\rm EOS}=0.05$
(bottom) along columns.  Some interesting features and trends are
apparent from Figure~\ref{fig:isomontage}.

First, for a relatively stiff EOS, $q_{\rm EOS}\simgt 0.3$, 
increasing the gas
fraction actually suppresses the amplitude of spiral structure in the
disk.  This is because in our full multiphase model, the effective
temperature of the gas is $T_{\rm eff} \simgt 10^5$ K over most of the
disk for $q=1$, and so the gas is dynamically hotter than the old
stars.  Increasing the gas fraction in this case makes the disk more
stable and less susceptible to amplifying non-axisymmetric patterns.
Thus, in the top row of Figure~\ref{fig:isomontage}, the spiral
structure which is apparent for $f_{\rm gas} = 0.1$ (top left) is
mostly washed out for $f_{\rm gas} = 0.99$ (top right).  This tendency
does not occur if the EOS is soft because then the effective
temperature is such that the gas is dynamically colder than the old
stars.  For $q_{\rm EOS} 
\le 0.125$ (bottom two rows), this results in unstable
behaviour if the gas fraction is large.

Second, the set of models shown in Figure~\ref{fig:isomontage} clearly
delineates stable cases from unstable ones.  For our full, stiff EOS
($q_{\rm EOS}=1$) 
even pure gas disks are stable (top row).  This is also true
for intermediate cases with $q_{\rm EOS}=0.5$.  However, if the effective
temperature of the gas is lower than that of the stars, very gas-rich
disks are unstable and fragment within a rotation period. Thus, for
example, with a nearly isothermal EOS ($q_{\rm EOS}=0.05$, bottom row), 
disks
with a gas fraction $f_{\rm gas} \simgt 0.3$ are already unstable
(bottom row).

\begin{figure*}
\begin{center}
\resizebox{7.5cm}{!}{\includegraphics{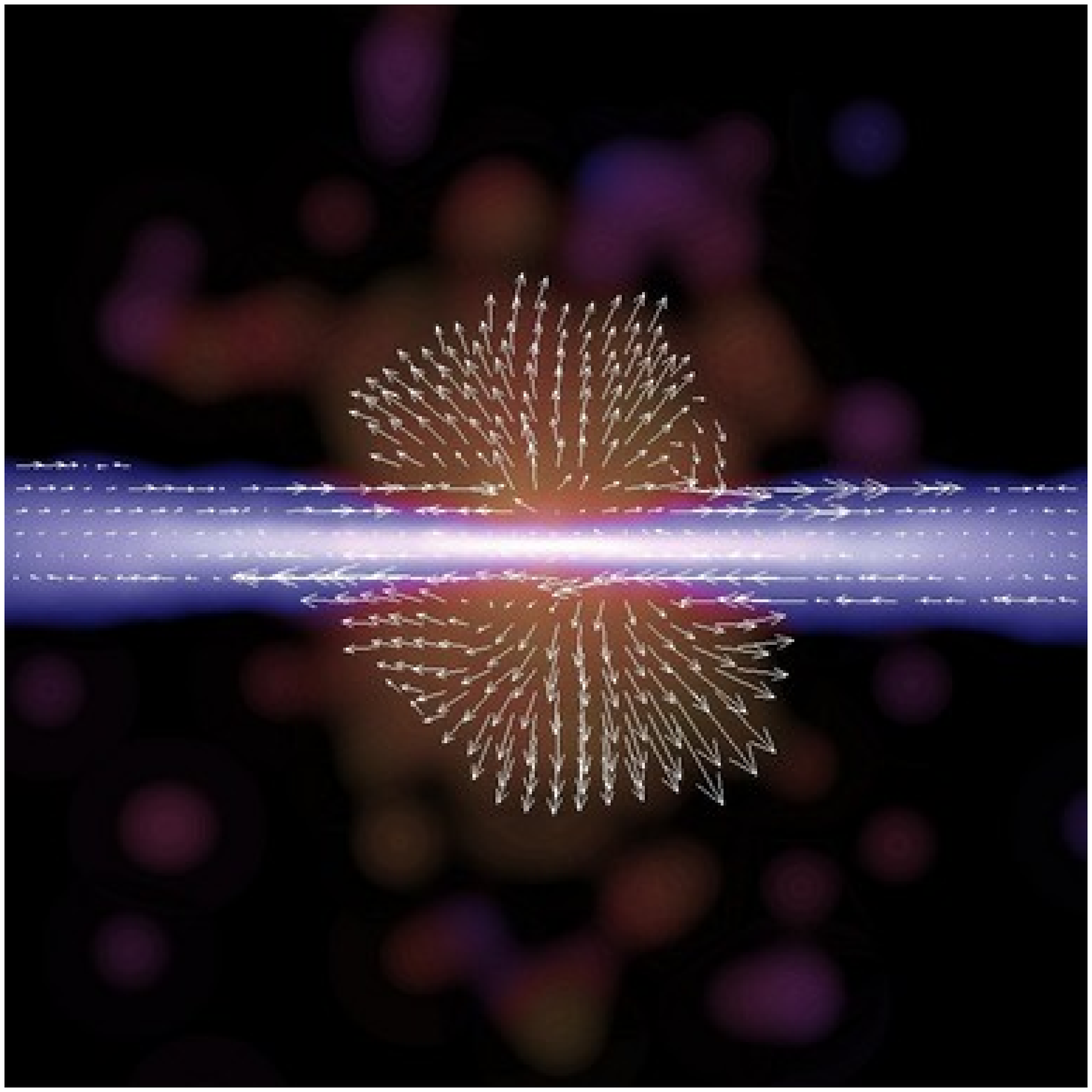}}\ %
\resizebox{7.5cm}{!}{\includegraphics{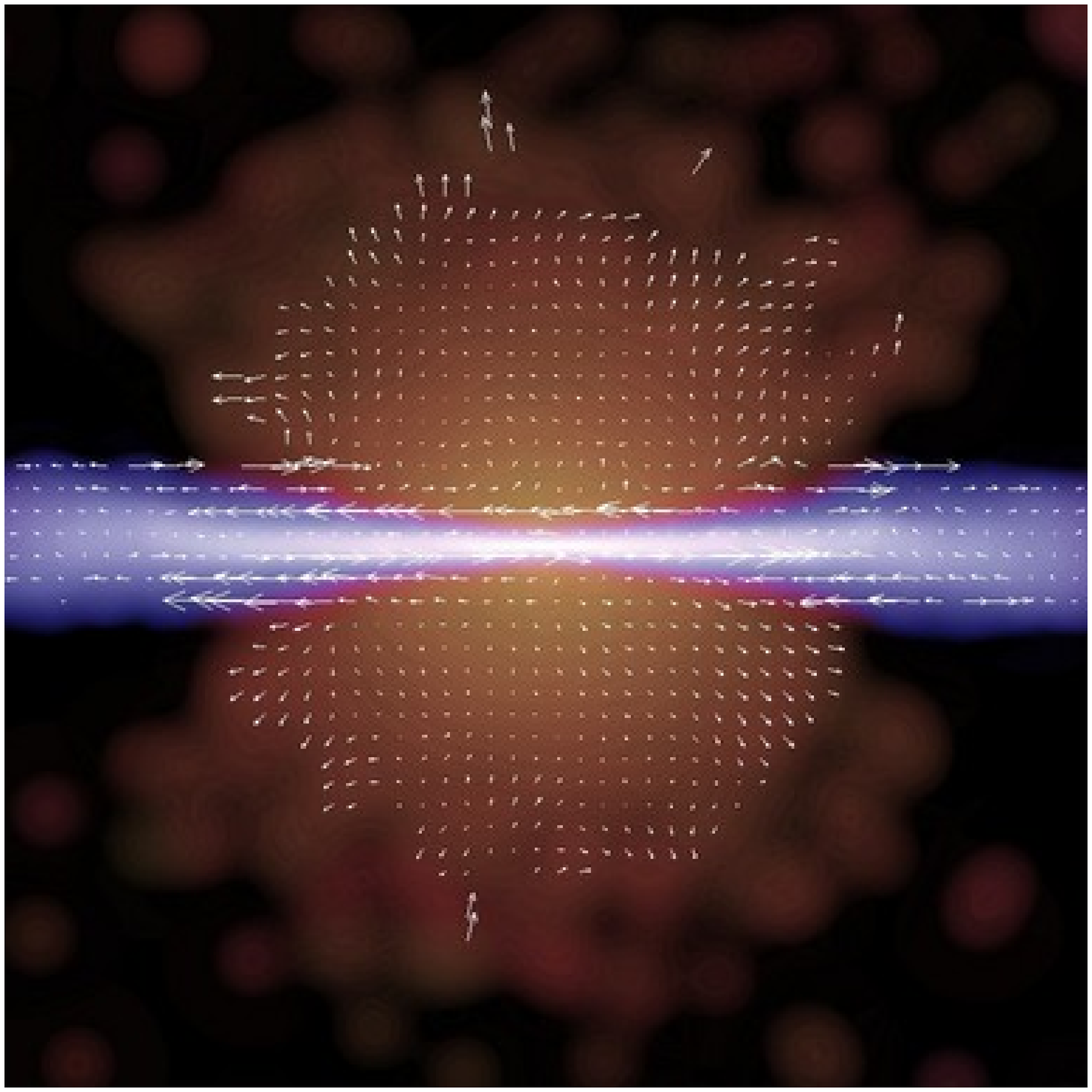}}
\end{center}
\caption{Edge-on view of the gas in an isolated disk galaxy with a
  bulge that hosts a growing, supermassive black hole in its nucleus.
  The two panels of side-length $32\,h^{-1}{\rm kpc}$ show snapshots
  at times $0.7\,{\rm Gyrs}$ (left) and $1.4\,{\rm Gyrs}$ (right)
  after the beginning of the simulation.  At the low accretion rates
  present in this quiescent galaxy, feedback energy from the black
  hole can nevertheless drive a weak wind into the halo.}
\label{fig:isobhwind}
\end{figure*}

The previous studies by \citet{Hernquist1995} and \citet{Mihos1996}
treated the ISM as an isothermal gas and included only weak feedback
in the form of a small input of random kinetic energy.  These authors
were unable to construct stable models with large gas fractions owing
to disk fragmentation like that shown in Figure~\ref{fig:isomontage}.
Supernova feedback in our multiphase model pressurises the gas making
the effective EOS stiffer, stabilising the disk.

A similar separation between stable and unstable behaviour is seen in
our galaxy models that include bulges.  We illustrate this in another
manner in Figures~\ref{fig:isosfrnob} and \ref{fig:isosfrb} where we
show the evolution of the global star formation rates in the disks for
models without and with bulges, respectively.

The behaviour in Figures~\ref{fig:isosfrnob} and \ref{fig:isosfrb}
clearly separates stable and unstable models.  In unstable disks,
fragmentation leads to a runaway growth in density perturbations,
yielding unsteady evolution in the global star formation rate. Thus,
all our galaxies are stable for stiff equations of state with $q_{\rm EOS}
\ge0.5$.  Models with a bulge are stable for all gas fractions with
$q_{\rm EOS}=0.25$, 
however those without a bulge (Figure~\ref{fig:isosfrnob})
are unstable with this EOS for $f_{\rm gas} \simgt 0.6$.  Softer
equations of state yield unstable behaviour for galaxies without
bulges for $f_{\rm gas} \simgt 0.4$.  For models with bulges,
instability sets in at $f_{\rm gas} > 0.8$ for $q_{\rm EOS}=0.125$ 
and $f_{\rm
gas} \simgt 0.6$ for $q_{\rm EOS}=0.05$.  The scalings with $q_{\rm EOS}$ 
and $f_{\rm
gas}$ are broadly consistent with the Toomre criterion.

\subsection{Models with black holes}
\label{modsbh}

We have also run models similar to those described in
Section~\ref{modsnobh}, but where we added a sink particle to
represent a black hole and allowed it to accrete gas from the galaxy,
increasing its mass and adding a source of feedback energy in addition
to supernovae.  The examples we discuss in this section are not
intended to provide a physical picture for the growth of supermassive
black holes in isolated galaxies because we have made no attempt to
tie the origin of the black holes to galaxy formation.  Nevertheless,
we can use these simulations to identify some generic features of our
description of the interaction between black holes and the ISM that
are relevant for dynamically evolving situations, like galaxy mergers.

The coupling between black hole accretion and surrounding gas has
several implications.  First, heating from AGN feedback energy can
drive outflows from galaxies if the accretion rate is sufficiently
high. This is shown in Figure~\ref{fig:isobhwind}, where the effect of
black hole feedback is illustrated for an isolated galaxy with a
bulge. For the low accretion rates relevant to the times indicated,
the deposition of thermal energy into the gas in the vicinity of the
black hole can drive a weak wind perpendicular to the plane of the
disk, as indicated by the velocity vectors in this figure.  These
outflows are reminiscent of those in our supernova-driven winds
(Figs. 5-8 in SH03).  More powerful outflows can be produced during
mergers when gas is driven into the centre of the remnant, fueling
strong nuclear accretion.

As implied in Section~\ref{bhole}, our description of black hole
accretion yields several distinct phases of evolution, depending on
the properties of the gas near the black hole.  For accretion rates
lower than Eddington, the black hole mass will evolve according to \be
M_{\rm BH} (t) \, = \, {{M_0}\over{1 \, - \, \chi \, M_0 \, t}} \, ,
\label{eq:bondevol}\ee where $M_0$ is the initial black hole mass and
\be \chi \, = \, {{4\pi \, \alpha G^2 \, \rho}\over{(c_s^2 +
v^2)^{3/2}}} \, .  \ee This result is valid only when the properties
of the gas near the black hole and the relative velocity do not evolve
significantly with time.  When the accretion rate is higher than
Eddington, the black hole grows exponentially with time, \be M_{\rm
BH} (t) \, = M_{\rm BH}(0) \, \exp \left ( {{t}\over{t_S}} \right )
\,, \label{eq:eddevol} \ee where the Salpeter time, \be t_S \, \equiv
\, {{\epsilon _r \, \sigma_T \, c}\over{4\pi \, G \, m_p}} \, ,  \ee
depends only on physical constants and on the radiative efficiency.
For a radiative efficiency of $10\%$, $t_S \approx 4.5 \times 10^7$
years.

In the absence of feedback effects, a low-mass black hole would grow
first according to equation (\ref{eq:bondevol}).  Once the black hole
becomes sufficiently massive, given the properties of the surrounding
gas, it then enters a period of exponential growth, described by
equation (\ref{eq:eddevol}). Note that the first, Bondi-limited growth
can be very slow if the initial black hole seed mass is small, or if a
small value for the coefficient $\alpha$ is selected. We typically
start our calculations with black hole seed masses of order
$10^5\,{\rm M}_\odot$ and set $\alpha$ large enough to allow a hole in
a gas-rich environment to reach the Eddington regime in about $\sim
0.5\,{\rm Gyr}$.

\begin{figure}
\begin{center}
\hspace*{-0.6cm}\resizebox{8.5cm}{!}{\includegraphics{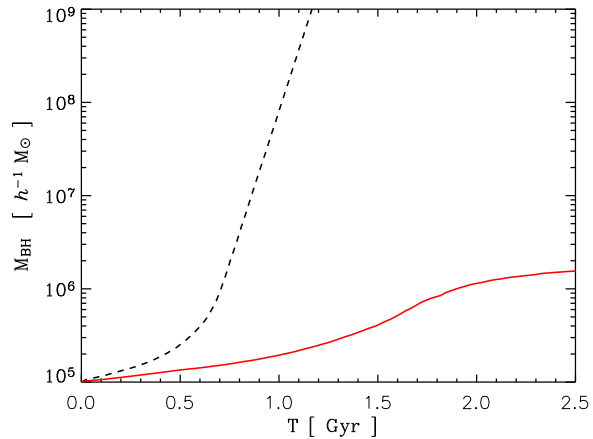}}
\end{center}
\caption{Evolution of the black hole mass in simulations of an
  isolated gas-rich galaxy. We show the black hole mass with (solid)
  and without (dashed) thermal feedback, respectively.}
\label{fig:isobhevol}
\end{figure}

We illustrate this behaviour in Figure~\ref{fig:isobhevol}, which
shows the growth of black holes in isolated, gas-rich galaxies, with
and without the impact of accretion feedback on the gas. If feedback
is neglected, the black hole grows in a manner consistent with the
expectations noted above, first according to equation
(\ref{eq:bondevol}) and later exponentially.  When the effects of
feedback are included, however, the growth is self-regulated and leads
to a saturated final state, that depends on the initial gas content of
the galaxy.  The self-regulation occurs because as the accretion rate
increases, the amount of energy available to affect nearby gas also
grows.  If this material is then heated and dispersed, the supply of
gas to grow the black hole will decline as accretion is shut off.

\section{Major Mergers}
\label{mmergers}

The approach we have adopted for handling various forms of feedback
makes it possible to explore physical effects that were not accessible
previously.  For example, \citet{Hernquist1995} and \citet{Mihos1996}
were able to model galaxies with only small gas fractions because
their treatment of feedback was insufficient to stabilise highly
gas-rich disks.  With our method, we can examine mergers between
galaxies with a very large gas content, as might be relevant to
objects at high redshifts.  Also, by including processes related to
black hole growth and accretion, we can study the interplay between
stellar and AGN feedback that is likely to be important for
understanding the nature of ultraluminous infrared galaxies (ULIRGs).

\begin{figure*}
\begin{center}
\resizebox{17.0cm}{!}{\includegraphics{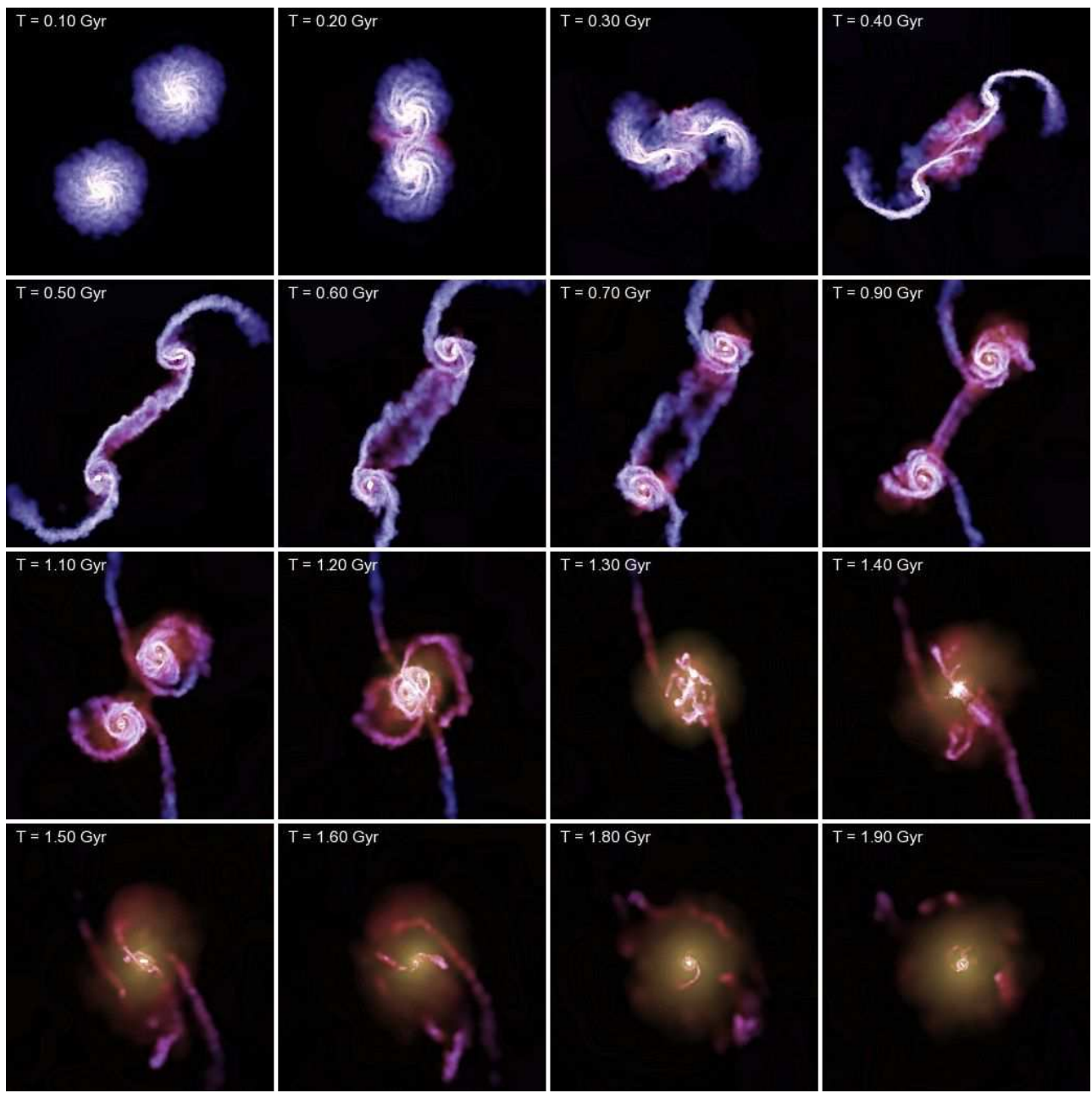}}
\end{center}
\caption{Evolution of the gas distribution in a major merger of two
  disk galaxies.  Colour hue encodes gas temperature, while brightness
  indicates gas density.  Each frame measures $50\,h^{-1}{\rm kpc}$ on
  a side, and the corresponding time of each image is given by the
  labels. In this simulation, a slightly softer EOS ($q_{\rm EOS}=0.25$)
  than in
  our default multiphase model was used.
\label{fig:mmtime}}
\end{figure*}

In what follows, we illustrate some of the consequences of our models
for stellar and AGN feedback by looking at major mergers of disk
galaxies.  A representative case is given in Figure~\ref{fig:mmtime},
which shows the distribution of gas during the merger of two
equal-mass disk galaxies from a prograde, parabolic orbit.  As the
disks pass by one another for the first time ($t=0.2 - 0.3$), strong
tidal forces yield extended tails and bridges.  Gas is shocked at the
interface between the two galaxies and the tidal response drives gas
into the central regions of the two galaxies.  When the galaxies
finally coalesce ($t=1.2 - 1.5$), much of the remaining gas is either
converted into stars during an intense burst, or shock-heated to
temperatures characteristic of the virial temperature of the remnant.

\subsection{Mergers without black holes}
\label{mmnobh}

\subsubsection{Comparison with previous work}

One of our goals is to see to what extent the strength of stellar
feedback influences starbursts during mergers.  As a starting point,
we compare our results to those obtained in earlier studies, in
particular, to the work of \citet{Hernquist1995} and
\citet{Mihos1996}.  These authors employed a highly simplified
treatment of star formation by modeling the ISM as an isothermal gas
with a temperature $T_{\rm ISM} = 10^4\,{\rm K}$, and including a weak
form of supernova feedback by injecting small amounts of kinetic
energy into the gas.  The amplitude of this feedback was adjusted so
that isolated disks forming stars quiescently had gas profiles similar
to those observed.

\begin{figure}
\begin{center}
\resizebox{8.0cm}{!}{\includegraphics{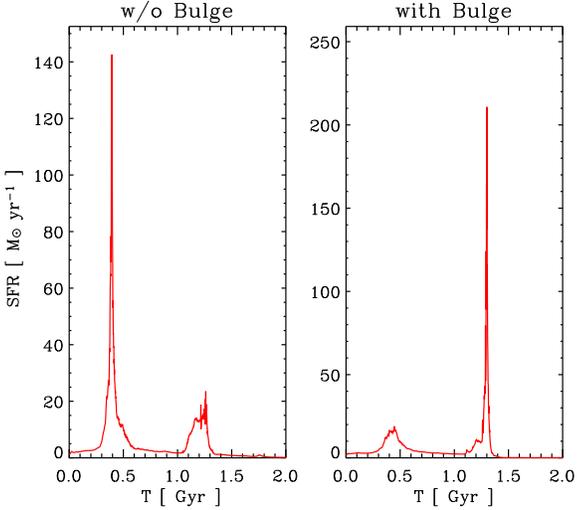}}
\end{center}
\caption{Evolution of the star formation rate in major mergers of disk
  galaxies, using a description of the ISM that mimics the approach of
  \citet{Hernquist1995} and \citet{Mihos1996}. We show results for
  models without (left) and with (right) bulges. The galaxies have the
  rotation curves shown in Figure~\ref{fig:galrot}, and the disks
  initially consisted of $10\%$ gas.}
\label{fig:mhsfr}
\end{figure}

We have modified our simulation code to mimic the star formation and
feedback algorithm of \citet{Mihos1994a} in an attempt to reproduce
their results. In Figure~\ref{fig:mhsfr}, we show the evolution of the
star formation rate in prograde mergers of equal-mass disk galaxies
with and without bulges.  We have selected the parameters and galaxy
models to match the results of \citet{Mihos1996} as closely as
possible.  The results in Figure~\ref{fig:mhsfr} can be compared
directly to e.g. Fig. 5a in \citet{Mihos1996}.

The morphologies of the merging galaxies roughly follows the time
sequence in Figure~\ref{fig:mmtime}, with only a modest dependence on
their structural properties.  When the galaxies first pass by one
another, relatively strong starbursts are triggered in each disk in
the case when bulges are not present.  Galaxies with bulges experience
much weaker starbursts during first passage.  \citet{Mihos1996} showed
that this difference arises from the relative ease with which the
disks can amplify $m=2$ disturbances in response to tidal forcing.
During final coalescence, a much strong starburst is excited in the
models with bulges because much of the gas in the bulgeless galaxies
has already been consumed by this time.  These trends agree well with
those identified by \citet{Mihos1996}.

The amplitudes of the starbursts shown in Figure~\ref{fig:mhsfr} are
comparable to, but somewhat stronger than those shown in Fig. 5a of
\citet{Mihos1996}, and considerably stronger than those found by
\citet{Barnes2004} (see e.g. his Fig. 5).  By varying the parameters
of our simulations, we have found that the evolution of the star
formation rate in mergers with only relatively weak feedback is
sensitive to numerical resolution.  If the star formation rate is tied
to gas density, the amplitudes of merger-induced starbursts depend on
the compressibility of the gas, which is influenced by both the
stiffness of the EOS, as well as dynamic range in resolution of the
numerical algorithm.  We have verified that differences in resolution
are mainly responsible for the residual discrepancies between the star
formation rates shown in Figure~\ref{fig:mhsfr} and those reported by
\citet{Mihos1996} and \citet{Barnes2004}.

\subsubsection{Multiphase treatment of ISM}
\label{mmmulti}

As a next step, we have performed a grid of merger simulations using
the multiphase treatment of supernova feedback described in
Section~\ref{starfeed}.  We have varied the structural properties of
the galaxies, the orbit of the mergers, and the parameters in our
multiphase model.  Below, we limit the discussion mainly to the
consequences of variations in the fraction of gas in each galaxy and
the strength of feedback to highlight the new types of outcomes that
are possible.

\begin{figure}
\begin{center}
\resizebox{8.0cm}{!}{\includegraphics{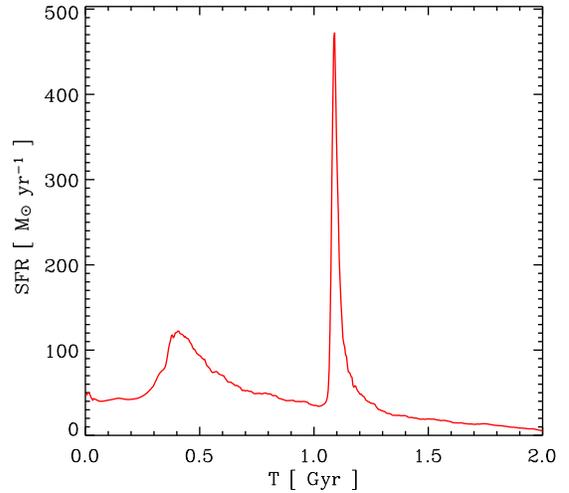}}
\end{center}
\caption{Evolution of the star formation rate in a major merger of two
  disk galaxies without bulges, and with disks that initially
  consisted of $99.9 \%$ gas.}
\label{fig:sfr88}
\end{figure}

In Figure~\ref{fig:sfr88}, we show the evolution of the star formation
rate from one of our simulations, a prograde, major merger of two disk
galaxies without bulges.  In this case, each disk initially consisted
of $f_{\rm gas} = 99.9\%$ gas and only $0.1\%$ of old stars.  The EOS
was relatively stiff with a softening parameter of $q_{\rm EOS}=0.5$.
(That is,
the EOS was linearly interpolated to lie midway between the curves in
Figure~\ref{fig:eosfig} for $\rho > \rho_{\rm th}$.)  Even with this
extreme gas content, the model disks are stable in isolation, owing to
the pressurisation of the gas from supernova feedback.  The parameters
of this simulation are deliberately chosen to be extreme to simplify
the discussion, but qualitatively similar results are obtained in more
general situations.

In the example shown in Figure~\ref{fig:sfr88} the star formation rate
peaks at roughly $500\, {\rm M}_\odot{\rm yr}^{-1}$, considerably
higher than in earlier simulations of equal-mass mergers of disks,
owing to the large gas content of each galaxy.  Star formation rates
at this level are similar to those inferred for systems at high
redshift such as Lyman-break galaxies and SCUBA sources, suggesting
that some of these objects may result from mergers of gas-rich disks.

Figure~\ref{fig:sfr88} also demonstrates that while the general
conclusions of \citet{Mihos1996} are obtained under broader
circumstances, some of their results clearly depend on the strength of
feedback from star formation.  In models with weak supernova feedback,
as in Figure~\ref{fig:mhsfr}, galaxies without bulges experience
stronger starbursts at first passage than during final coalescence.
The evolution shown in Figure~\ref{fig:sfr88} is rather different,
with the strongest starburst accompanying the final merger.  While the
merging galaxies in this example still develop a bar-like structure in
response to tidal forcing, the gas is sufficiently hot dynamically
that it is not strongly compressed during this phase of evolution,
limiting the strength of the initial starburst.  The more intense
starburst during the final merger results from strong shock
compression of the gas and gravitational torquing when the galaxies
coalesce, in a manner reminiscent of the mergers with bulges examined
by \citet{Mihos1996, BarnesHernquist1991, BarnesHernquist1996}.

An interesting aspect of the simulation used in Figure~\ref{fig:sfr88}
concerns the structure of the merger remnant. A considerable amount of
gas is left behind following the merger. Much of it settles into an
extended disk, surrounding a relatively compact, rapidly rotating
remnant of newly formed stars.  Overall, the remnant is
morphologically closer to a spiral galaxy with a bulge than to an
elliptical. Further implications and a detailed analysis of this
simulation are presented in \citet{Springel2004b}.

\subsection{Mergers with black holes}
\label{mmbh}

An even greater diversity of outcomes results during a merger if the
impact of central, supermassive black holes is included in addition to
feedback from star formation.  As we discuss in \cite{DiMatteo2004}
and \citet{Springel2004}, mergers of galaxies that involve star
formation, supernova feedback, black hole growth and accretion, and
AGN feedback yield a generic time history in which the tidal
interaction drives gas into the centre of the remnant, fueling a
starburst and triggering the rapid growth of the central black
hole(s). Feedback from accretion onto the black hole(s) influences the
thermodynamic properties of the surrounding gas, eventually expelling
it from the inner regions of the remnant, terminating the episode of
rapid star formation and self-limiting the growth of the black
hole(s). The starburst and AGN activity are coeval, but offset in time
owing to the detailed form of the response of the gas to the
feedback. Thus, while we expect starbursts and AGN activity to be
correlated in this picture, the remnant will be primarily seen in
different stages as the merger progresses and will evolve from one
type of object into the other, depending on the observed wavelength
regime.

\begin{figure*}
\begin{center}
\resizebox{4.8cm}{!}{\includegraphics{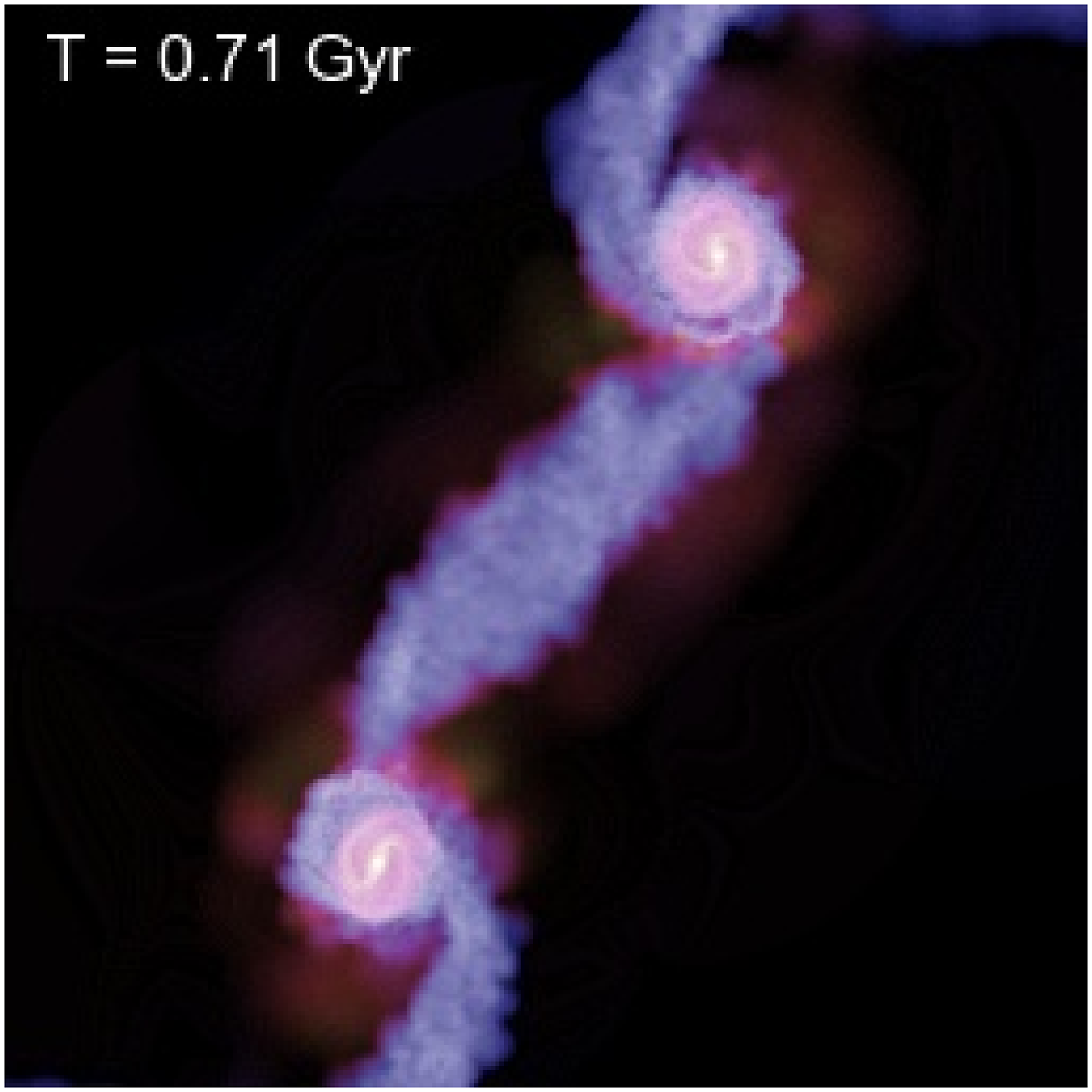}} %
\resizebox{4.8cm}{!}{\includegraphics{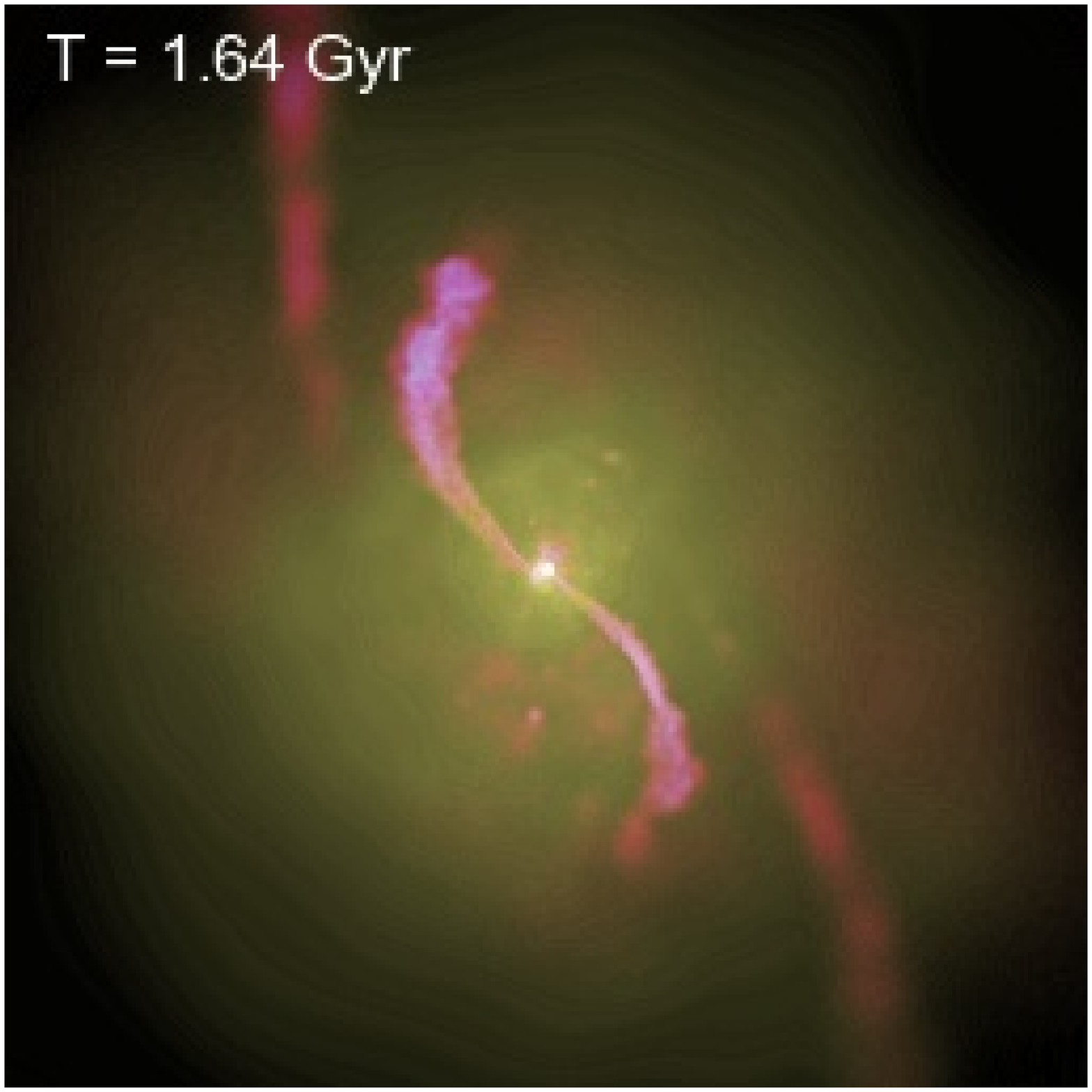}} %
\resizebox{4.8cm}{!}{\includegraphics{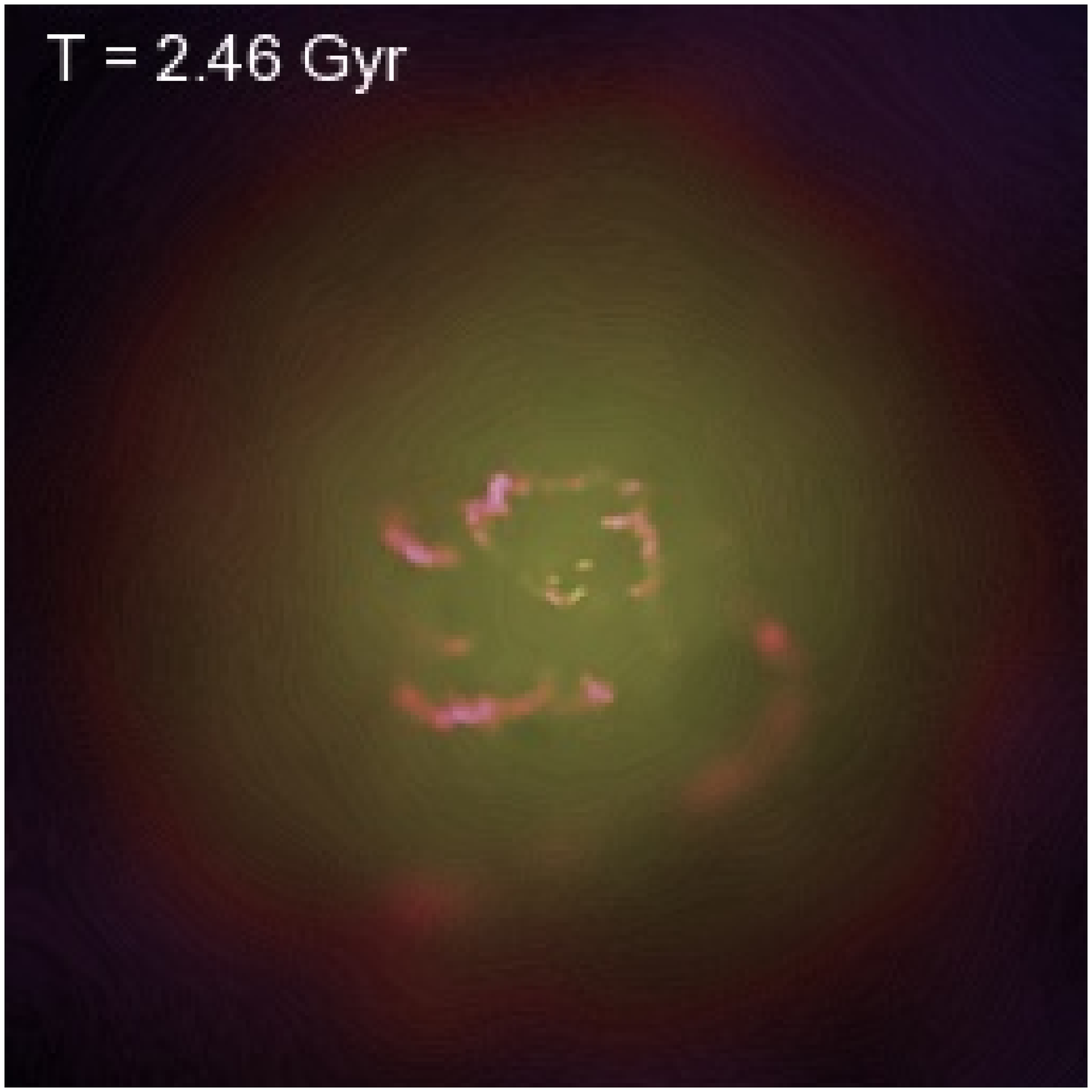}}\\
\resizebox{14cm}{!}{\includegraphics{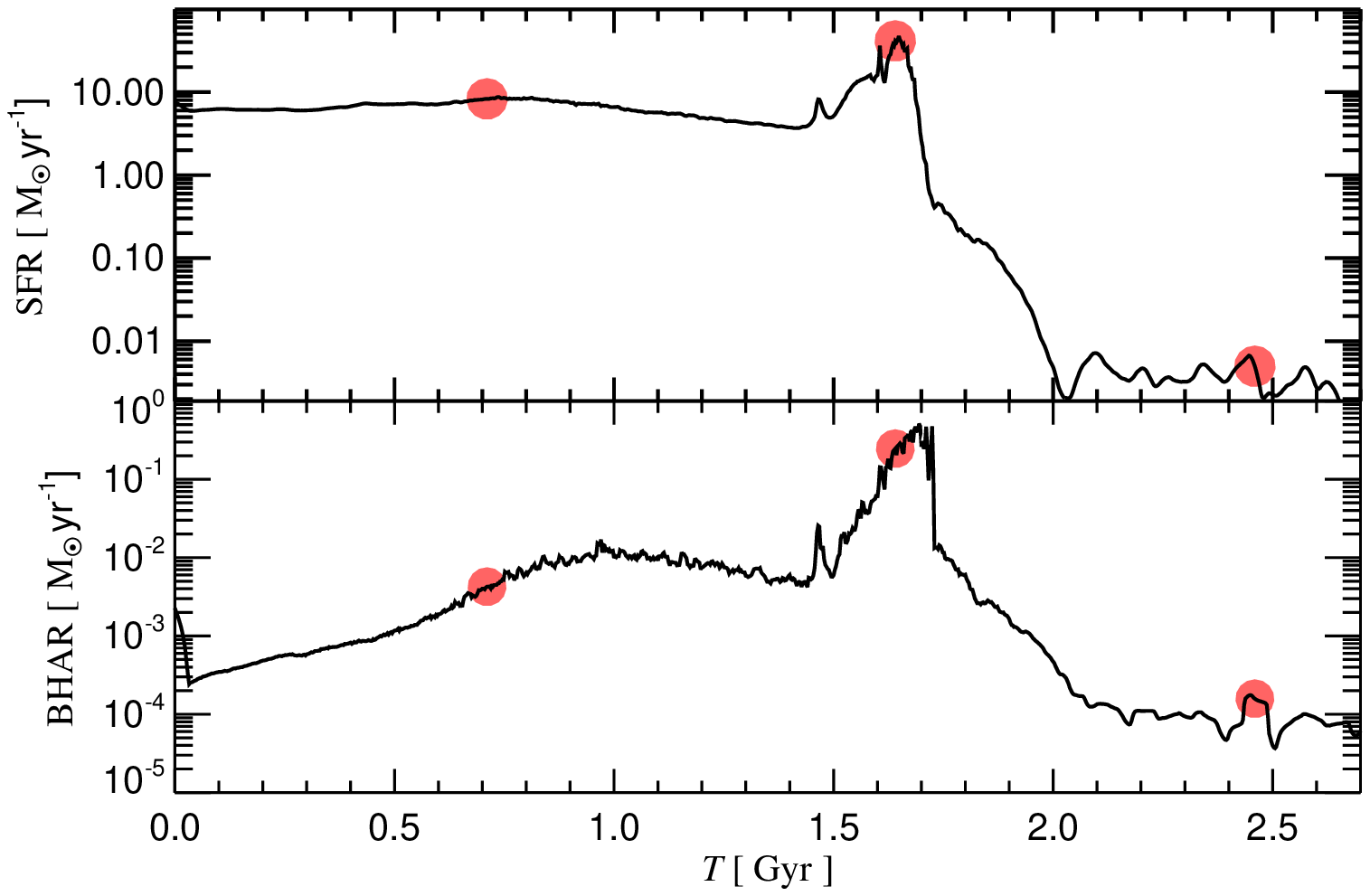}}\vspace*{-0.5cm}\\
\end{center}
\caption{Merger of two disk galaxies, including the effects of black
  hole growth and AGN feedback.  The images show the gas distribution
  in the two disks at three different times, where colour hue encodes
  temperature while brightness measures gas density. The bottom panels
  show the time evolution of the accretion rate onto the black holes
  (top) and the star formation rate (bottom).  The red symbols in
  these panels mark the three times shown in the images on top.  The
  first snapshot shows the system just after the first passage of the
  two disks. The second snapshot captures the system when the galaxies
  are coalescing, at which point the star formation and accretion
  rates peak. Finally, the third snapshot shows the system after the
  galaxies have fully merged, and most of the mass has settled into a
  slowly evolving remnant.}
\label{fig:sfrbha}
\end{figure*}

In Figure~\ref{fig:sfrbha} we show three snapshots from a simulation
that includes our model for the growth of black holes.  In this
example, the galaxies included bulges and the disks were $10\%$ gas.
Following the first passage (shown on the left), but before the
galaxies coalesce, the disks are distorted by their mutual tidal
interaction, but only a relatively weak starburst is triggered,
because the bulges stabilise the disks against a strong $m=2$
response.  The black holes do not accrete significantly during this
phase of evolution, and so neither galaxy would be observable as an
AGN.  When the galaxies merge but before the gas has been consumed
(middle), tidal forces trigger a nuclear starburst and fuel rapid
growth of the black holes.  The peak in the starburst occurs near the
time indicated in Figure~\ref{fig:sfrbha}, but the peak in the black
hole growth is offset, owing to the delayed action of AGN feedback on
the gas.  In principle, the system should now be both a starburst and
an AGN, but it is likely that the quasar-activity would be obscured by
surrounding gas and dust, at least during the beginning of the burst.
Only during the final stages of evolution could the remnant be visible
as an AGN, when outflows remove the dense layers of gas around it.

As the remnant settles into a relaxed state, star formation and black
hole accretion are rapidly quenched by the expulsion of gas from the
centre, as a consequence of AGN feedback.  The remnant would no longer
be observable as either a starburst or an AGN and will age rapidly,
quickly resembling an evolved red galaxy \citep{Springel2004}. At this
point the black hole growth ceases and the black hole mass attained is
consistent with those expected from the observed $M_{\rm BH} - \sigma$
relation \citep{DiMatteo2004}.

\begin{figure}
\resizebox{8.5cm}{!}{\includegraphics{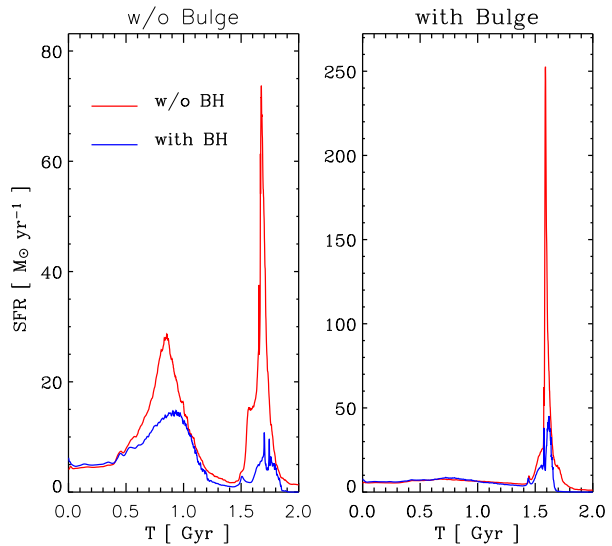}}
\caption{Star formation rates in mergers of galaxies without (left)
and with (right) bulges. In each panel, the two curves show the
outcome with and without black holes.}
\label{fig:sfrbhnobh}
\end{figure}

The interplay between black hole growth and the physics of the ISM in
simulations like that shown in Figure~\ref{fig:sfrbha} has significant
implications for the observed properties of these systems.  In
Figure~\ref{fig:sfrbhnobh}, we show the evolution of the intrinsic
star formation rates in simulations of galaxy mergers with and without
bulges and including or neglecting central black holes.  This figure
shows that the presence of the black holes and the feedback energy
derived from their growth can dramatically alter the nature of the
starburst resulting from a merger. The peak amplitude of the starburst
in the merger which included bulges is lowered by about a factor of
five, owing to AGN feedback. We note, however, that the observed
emission from such an object would include both the radiation from the
starburst as well as the energy output from black hole accretion.  If
most of the AGN emission is reprocessed into optical or infrared
radiation by surrounding gas and dust before it escapes from the inner
regions of the remnant, it could be incorrectly attributed to the
starburst, rather than being associated with a buried quasar.  A more
complete understanding of the implications of these results will
require a detailed modeling of the spectral energy distribution from
the combination of the starburst and the obscured AGN.

\begin{figure*}
\begin{center}
\vspace*{-0.3cm}\resizebox{15.0cm}{11cm}{\includegraphics{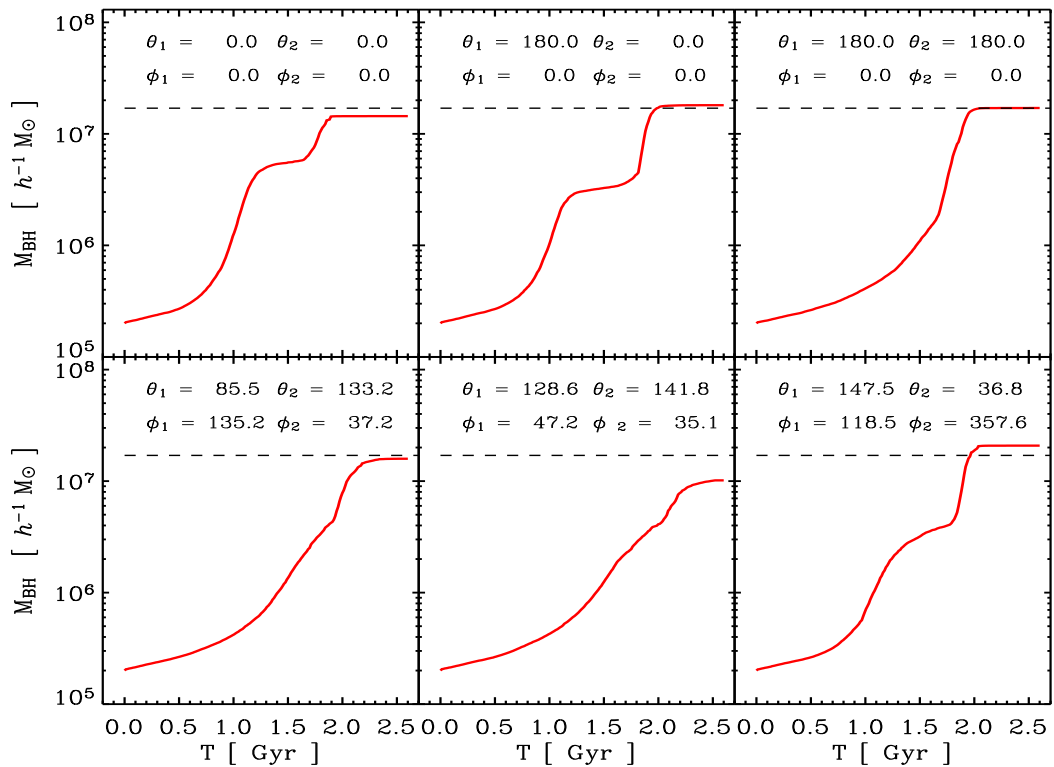}}\vspace*{-0.5cm}
\end{center}
\caption{Evolution of the total black hole mass in merger simulations
  that differ in their encounter geometries. The top three panels show
  results where the disks are oriented in the orbital plane, while the
  bottom three panels are for simulations with random disk
  orientations. The angles $(\theta, \phi)$ specify the orientation of
  the spin vector of one of the disk galaxies relative to the orbital
  plane \citep[see][for a sketch of the orbital
  configuration]{Duc2000}.}
\label{fig:bhgrowth}
\end{figure*}

We have varied the orbital geometry and structural properties of the
merging galaxies to see if the results described above are generic.
While the detailed outcome does depend on e.g. the orbit, many of the
overall properties are relatively unaffected.  An example is shown in
Figure~\ref{fig:bhgrowth} where the evolution of the total black hole
mass is given for six different orbital configurations.  The top three
panels show prograde-prograde, prograde-retrograde, and
retrograde-retrograde mergers (left to right), while the bottom three
panels show cases where the disk orientations were chosen randomly.
While the detailed response varies from model to model, the final
total black hole mass is insensitive to the orbital parameters.  In
all cases, the final black hole mass is $\sim 2-3 \times 10^7\, {\rm
M}_\odot$.  This value is consistent with the observed correlation
between black hole and spheroid mass for the remnants produced in
these mergers, as shown in more detail in \citet{DiMatteo2004}. The
simulation results are in accord with the view that black hole growth
is regulated by AGN feedback and the dynamical response of gas to this
supply of energy.

We note that in our model the AGN feedback energy is deposited
isotropically in the gas around the black hole. While this appears to
be a natural approximation given that we cannot resolve the detailed
accretion physics close to the hole, observations show that black hole
outflows are typically bi-polar (although not necessarily strongly
collimated) and often result from powerful jets. It is possible that
our isotropic coupling enhances the effects of the feedback, but this
effect should be compensated in part by our comparatively conservative
assumption for $\epsilon_{\rm f}$. We also caution that the physical
nature of the coupling of AGN feedback to the surrounding gas is not
fully understood yet. In particular, there may be important feedback
mechanisms other than thermal heating.  For example, radiation
pressure may play a very important role as well \citep{Murray2004}.

The presence of central black holes can also have a significant impact
on the stellar density structure of merger remnants.
\citet{Mihos1994b} found that their remnants typically had centrally
peaked luminosity profiles from the compact stellar cores left behind
by starbursts.  In our new simulations, the remnants have smoother
luminosity profiles into their centres, in much better agreement with
observations of elliptical galaxies (Springel, Di~Matteo \& Hernquist,
2004, in preparation). This outcome is driven by a combination of the
consumption of dense gas by the black holes, the dynamical heating of
the gas as the black holes spiral together and merge, and the
dispersal of gas by AGN feedback. Our results thus indicate that black
hole growth may be an integral part of spheroid formation.

\section{Conclusions}
\label{concs}

In this paper, we have introduced a new methodology for simultaneously
modeling star formation and black hole accretion in hydrodynamical
simulations of isolated and merging galaxies. Our approach uses the
concept of forming ``macroscopic'', coarse-grained representations of
sub-resolution physics, allowing us to study the effects of these
processes on larger, resolved scales.  To illustrate the principal
effects of our methods and of the physics we describe, we have focused
on simulations of isolated and merging galaxies. Throughout, we have
concentrated on the methodological aspects of our work, rather than on
an in-depth analysis of the physical results of our simulations. We
provide the latter in related papers
\citep[e.g.][]{DiMatteo2004,Springel2004}.

For our simulations, we have constructed compound galaxy models that
consist of a dark matter halo, a rotationally supported disk of gas
and stars, and a central bulge, with independent parameters describing
each of the structural components. Our approach improves upon previous
methods by accounting for gas pressure forces, as well as featuring a
flexible numerical approach for computing the gravitational potential
for non-trivial, radially varying mass distributions in the disk. As a
result, we are able to construct very stable disk galaxy models even
for high gas fractions.

In simulations of isolated galaxies, we have emphasised the importance
of the treatment of feedback processes in the dense star-forming ISM
for the stability of disk galaxies. Simulations that use a simple
isothermal equation of state, or a weak form of kinetic feedback, have
gas disks that are very susceptible to axisymmetric perturbations. As
a result, galaxies with large gas surface mass density cannot be
evolved in a stable fashion for a long time in such models. Our
multiphase model for the ISM encapsulates the effects of local
supernova feedback in the form of an EOS, which is comparatively
stiff. When this description for the ISM is chosen, the dense gas is
stabilised by supernovae pressurising the gas, so that much larger gas
fractions than in previous studies become possible.

We used simulations of isolated galaxy models to examine the
properties of the various phases of accretion possible in our model
for the growth of black holes.  In particular, we showed how the black
hole growth of a small seed black hole accelerates in the Bondi regime
and eventually reaches Eddington-limited, exponential growth. However,
the growth process can be slowed down and regulated by the feedback
energy associated with the accretion. We assumed a simple thermal
coupling of a small fraction of the hole's bolometric luminosity to
the surrounding gas; this can increase the local sound speed of the
gas and thereby reduce the Bondi accretion rate, or in extreme cases
of high accretion, generate a powerful, pressure-driven quasar wind.

Using a number of simulations of major galaxy mergers, both with and
without black holes, we have explored a few basic consequences of our
methods.  When we model the star-forming gas with an isothermal
equation of state and weak feedback, our results are in good agreement
with previous work by \citet{Mihos1996}. However, our multiphase
treatment of the ISM allows us to investigate galaxy mergers with much
higher gas fractions than possible before. Here, new types of outcomes
are possible for major mergers.  For example, very high star
formation rates of several hundred solar masses per year can be
reached in gas-rich mergers.  In merger simulations with black holes
we have shown that the presence of accreting black holes can
dramatically alter the dynamics of the merger.  The feedback energy
associated with the growth of the hole owing to the tidally triggered
inflow of gas has a direct effect on the strength of the nuclear
starburst. During the final galaxy coalescence, the BH can expel the
gas from the centre in a powerful outflow, quenching the starburst on
a short timescale. The resulting elliptical remnant is then gas-poor
and reddens quickly.

Already, these exploratory results show clearly that adding an
accreting supermassive black hole in the centre can have a decisive
influence on galaxies.  In fact, we think this indicates that the
build-up of supermassive black holes and massive galaxies are an
intertwined process which requires a self-consistent treatment to be
meaningfully addressed.  Our new numerical methodology represents the
first attempt to do carry out this demanding task in cosmological
hydrodynamical simulations, opening up the possibility to jointly
study the formation of galaxies and the build-up of supermassive black
holes hosted by them. Our methods thus provide a powerful tool to gain
new insights into the process of hierarchical galaxy formation.

\section*{Acknowledgements}
This work was supported in part by NSF grants ACI 96-19019, AST
00-71019, AST 02-06299, and AST 03-07690, and NASA ATP grants
NAG5-12140, NAG5-13292, and NAG5-13381. The simulations were performed
at the Center for Parallel Astrophysical Computing at
Harvard-Smithsonian Center for Astrophysics.

\bibliographystyle{mnras}
\bibliography{paper}

\end{document}